\documentclass[fleqn,10pt]{wlscirep}
\usepackage[utf8]{inputenc}
\usepackage[T1]{fontenc}
\usepackage[english]{babel}
\usepackage[caption=false]{subfig}

\usepackage{float}
\usepackage{mathtools}

\title{The geography of innovation dynamics}

\author[1,2,3,*]{Matteo Straccamore}
\author[2,4,1,3]{Vittorio Loreto}
\author[2,4,1]{Pietro Gravino}

\affil[1]{Centro Ricerche Enrico Fermi, Via Panisperna 89/A, 00184, Rome, Italy.}
\affil[2]{Sony CSL Paris Research, 6, Rue Amyot, 75005, Paris, France.}
\affil[3]{Sapienza Univ. of Rome, Physics Dept., Piazzale Aldo Moro 2, 00185, Rome, Italy.}
\affil[4]{Sony CSL Rome Research, Joint Initiative CREF-Sony, Centro Ricerche Enrico Fermi, Via Panisperna 89/A, 00184, Rome, Italy.}

\affil[*]{matteo.straccamore@uniroma1.it}

\begin{abstract}
Cities and metropolitan areas are major drivers of creativity and innovation in all possible sectors: scientific, technological, social, artistic, etc. The critical concentration and proximity of diverse mindsets and opportunities, supported by efficient infrastructures, enable new technologies and ideas to emerge, thrive, and trigger further innovation. Though this pattern seems well established, geography's role in the emergence and diffusion of new technologies still needs to be clarified. An additional important question concerns the identification of the innovation pathways of metropolitan areas. Here, we explore the factors that influence the spread of technology among metropolitan areas worldwide and how geography and political borders impact this process. Our evidence suggests that political geography has been highly important for the diffusion of innovation till around two decades ago, slowly declining afterwards in favour of a more global innovation ecosystem. Further, the visualisation of the evolution of countries and metropolitan areas in a 2d space of competitiveness and diversification reveals the existence of two main innovation pathways, discriminating between different strategies towards progress. Our work provides insights for policymakers seeking to promote economic growth and technological advancement through tailored investments in prioritarian innovation areas.

\end{abstract}
\begin{document}
\flushbottom
\maketitle

\section{Introduction}
In our increasingly interconnected world, diffusion processes play a crucial role in determining the evolution of our societies. For this reason, a well-established and growing literature is focusing on studying the different instances of the phenomenon, from information diffusion in social networks~\cite{colbaugh2012early,kim2012discovery} to the spreading of diseases~\cite{brockmann2013hidden,melo2021heterogeneous,mazzoli2023spatial}. Particular attention converged on the diffusion of innovations~\cite{weil2018diffusion,lengyel2020role} and technologies~\cite{geroski2000models,comin2010exploration,comin2006five}. The adoption of patent data to monitor technological innovation is well established~\cite{frietsch2010value,griliches1998patents,leydesdorff2015patents}. For the past few decades, patent data have become a workhorse for the literature on technical change, due mainly to the growing availability of data about patent documents~\cite{youn2015invention}. This ever-increasing data availability (e.g., PATSTAT, REGPAT and Google Patents~\cite{hall2001nber}) has facilitated and prompted researchers worldwide to investigate various questions regarding the patenting activity. For example, on the nature of inventions, their network structure, and their role in explaining technological change~\cite{strumsky2011measuring,strumsky2012using,youn2015invention}.
One of the characteristics of patent documents is the presence of codes associated with the claims in patent applications. These codes mark the boundaries of the commercial exclusion rights demanded by inventors. Claims are classified based on the technological areas they impact, according to existing classifications (e.g., the IPC classification~\cite{fall2003automated}), to allow the evaluation by patent offices. Mapping claims to classification codes allows localizing patents and patent applications within the technology ``semantic'' space~\cite{jun2011ipc}.\\
In addition to the semantic space defined through technological codes, patents and innovations live in a physical space. It is known, for instance, the role that cities and metropolitan areas play in fostering creativity and innovation. Thanks to a critical concentration and proximity of diverse mindsets and opportunities, urban infrastructures enable new technologies and ideas to emerge, thrive, and trigger further innovation. Still, more is needed to know about the interplay between geography's role and the innovation processes' semantics. Technology and innovation diffusion processes take place, in fact, in a geographical layer that still needs to be studied, both from the physical and political points of view.

%SPIEGARE IMPORTANZA CITTà E PERCHE LE SCEGLIAMO, poi continuare con:
%Cities are essential for economic growth and development. According to a report by the World Bank~\cite{bank2018world}, cities generate about 80\% of global GDP. They attract businesses and industries, creating jobs and driving innovation~\cite{glaeser2012triumph}. In addition, cities provide a concentration of services such as healthcare, education, and cultural amenities, which are essential for a high quality of life~\cite{florida2017new}. They also foster social interaction and networking, allowing for the exchange of ideas and the development of new knowledge~\cite{jacobs1961jane}. From an environmental perspective, cities can be more sustainable than rural areas due to their greater efficiency in resource use and transportation~\cite{newman2015end}. Finally, cities are crucial for political organization, serving as the seats of government and allowing for efficient administration and governance of large populations~\cite{sassen2013global}. Overall, cities play a vital role in shaping our society and are essential for both economic and social development.

Cities and metropolitan areas (MAs) appear thus as the right level to investigate the role of geography in innovation processes. To date, approximately 55\% of the global population lives in urban areas, which represent the core of innovation~\cite{florida2017city,boschma2015relatedness}, economy~\cite{jacobs2016economy}, science~\cite{leydesdorff2010mapping}, and much more. According to a report by the World Bank~\cite{bank2018world}, MAs generate about 80\% of global GDP. They attract businesses and industries, creating jobs and driving innovation~\cite{glaeser2012triumph}; also, from an environmental perspective, MAs can be more sustainable than rural areas due to their greater efficiency in resource use and transportation~\cite{newman2015end}. For all these reasons, we focus on metropolitan areas as the smallest geographical entities, after countries and regions, essential for economic growth and development. 

Many recent studies have relied on network-based techniques to unfold the complex interplay among patents, technological codes, and geographical reference areas. We decided to use the framework of bipartite networks~\cite{asratian1998bipartite}, which are suitable whenever systems involve interactions between pairs of entities. For example, in ecology, interactions between two types of species can be described using bipartite networks, such as plant-pollinator networks~\cite{lopezaraiza2007impact} or seed-disperser networks~\cite{fedriani2014hierarchical}. Bipartite networks are also used in social~\cite{koskinen2012modelling}, economic~\cite{straccamore2022will,tacchella2012new}, and biological~\cite{pavlopoulos2018bipartite} systems.

%Story telling
With the tools described above and a specific focus on metropolitan areas, this paper investigates the factors that influence the spread of technology among metropolitan areas worldwide and how geography and political borders impact this process. We reveal that the current innovation pathways can be effectively predicted if one considers a non-trivial interplay between, on the one hand, the similarity between the technological content of cities and, crucially, belonging to the same country. In particular, our evidence suggests that political geography has been highly important for the diffusion of innovation till around two decades ago, slowly declining afterwards in favour of a more global innovation ecosystem. To this end, we improved current similarity-based prediction algorithms, i.e., algorithms based on the principle that the more two MAs are technologically similar, the higher the probability they will accomplish similar evolutionary technological paths. In particular, the improvement is substantial to forecast the so-called MAs technical "debut", i.e., the first-ever patent produced by a MA with a given technological code, where current models cannot formulate predictions.

We further visualise the evolution of countries and metropolitan in a 2d space of competitiveness and diversification. To this end, we adopted the UMAP dimensionality reduction algorithm~\cite{mcinnes2018umap} to visualise the different technological paths of countries and MAs. We discover the existence of two main innovation pathways, discriminating between different strategies towards progress. For instance, "Western'' countries and BRICS-like countries follow very different routes in this space, which we can define in terms of distinctive technological traits.

The paper is organised as follows. Section~\ref{sec:data} describes the data used in this work. In Section~\ref{sec:methods}, we introduce the methodologies used in our work, explaining the details of the similarity measures and testing procedures adopted. In Section~\ref{sec:results}, we present the results discussing the relevance of political geography, i.e., belonging to the same country, to obtain better predictive results, in particular, to predict the emergence of a brand-new technology in the portfolio of a given MA. We also display the innovation pathways of countries and MAs. Finally, in Section~\ref{sec:discussion}, we summarise the main results and highlight the hints the present work can give to future works addressing the questions arising from this study.

%%%%%%%%%%%%%%%%%%%%%%%%%%%%%%%%%%%%%%%%%%
\section{Data}\label{sec:data}
\subsection*{Technology Codes and Metropolitan Areas (MAs)}
We adopt the PATSTAT database (\url{www.epo.org/searching-for-patents/business/patstat}) to provide information about patents and technology codes. The database contains approximately 100 million patents registered in about 100 Patent Offices. Each patent is associated with a code that uniquely identifies the patent and a certain number of associated technology codes. The WIPO (World International Patent Office) uses the IPC (International Patent Classification) standard~\cite{fall2003automated} to assign technology codes to each patent. IPC codes make a hierarchical classification based on six levels called digits, which give progressively more details about the technology used. The first digit represents the macro category. For instance, the code Cxxxxx corresponds to the macro category "Chemistry; Metallurgy" and Hxxxxx to the macro category "Electricity". Considering the subsequent digits, we have, for instance, with C01xxx, the class "Inorganic Chemistry" and with C07xxx the class "Organic Chemistry".\\
For the metropolitan areas (MAs), we adopted a database (see next section) to match the unique patent identifier and its technology code to the corresponding MA. To geolocalise the patents, we adopted the De Rassenfosse et al. database~\cite{de2019geocoding} that contains entries on 18.9 million patents from 1980 to 2014. This is the first dataset about first filing applications from around the world, organised according to the location of applicants, i.e., companies or laboratories. This information helps study the geography of innovation and understand the spatial distribution of patented inventions. The geolocalisation is performed by linking the postal codes of applicant addresses to latitude and longitude and, as a result, to countries, regions, and MAs. The database contains information about the first application and assigns multiple technology codes to patents with more than one. The data is sourced from PATSTAT, WIPO, REGPAT, and the Japanese, Chinese, German, French, and British patent offices. Finally, each patent has unique identifiers, technology codes, and geographical coordinates (latitude and longitude). More information about De Rassenfosse et al. and PATSTAT database can be found in the Supplementary Information.

\subsection*{Data Preparation}
To clean the data, the first step consists of associating the technology codes of a patent with a specific MA by matching latitude and longitude information for each patent with the MAs borders obtained by the Global Human Settlement Layer~\cite{Schiavina2019GHS}. This way, we can select the patents within each MA's boundaries with their technology codes. Once this operation is completed, it is possible to build, year by year, the bipartite network that links MAs to technology codes. We represent the bipartite networks through bi-adjacency rectangular matrices $\textbf{V}^y$ whose elements $V_{a,t}^y$ are integers indicating how many times a technology code $t$ appeared in different patents in a given MA $a$ in year $y$. \\
Our network features $2865$ MAs connected to $650$ 4-digit technology codes. We decided to work with four digits instead of more or less because with the 4-digit we can have a technological resolution such that these are neither too similar nor too far apart. With more digits, we would have trivial results: for example, the 4-digit code A01C (Planting; Sowing; Fertilising) contains codes A01C-15 (Fertiliser distributors) and A01C-21 (Methods of fertilising). With fewer digits, we would have the opposite problem. In addition, multiple digits would have inherent problems with the PATSTAT database due to changes in database versions. Over time, new codes are born, or others are removed. The 4-digits choice appears as the most stable.\\
Our networks are represented by a set of matrices $\textbf{V}^y$ for each year, $y$, from $1980$ to $2010$. Each year $y$ matrix element $V_{at}^y$ counts how many times, in the year $y$, the technology $t$ appears in the MA $a$. Finally, we binarise the matrices $\textbf{V}$ simply using $0$ as a threshold to obtain $30$ $\textbf{M}^y$ matrices:
\begin{equation*}
    M_{at}^y = \begin{cases} 1\ \ \ \text{if}\ \ \ V_{at}^y \ne 0\\ 0\ \ \ \text{if}\ \ \ V_{at}^y = 0 \end{cases}
\end{equation*}
We decided to apply this binarisation procedure instead of the standard approaches like Revealed Comparative Advantage (RCA)~\cite{balassa1965trade} because we are interested to know which MA is adopting for the first time a given technology.
%%%%%%%%%%%%%%%%%%%%%%%%%%%%%%%%%%%%%%%%%%
\section{Methods}\label{sec:methods}

\subsection*{Similarity measures}
By the term \textit{Similarity}, we mean a measure of closeness between nodes in the same layer. In previous studies~\cite{hidalgo2007product,albora2021product,tacchella2021relatedness}, the similarity in the layer of items was used to study how an element of the layer of users may evolve in the future. For example, in~\cite{straccamore2022will}, the similarity between technologies was used to predict the future technology production of firms. In~\cite{albora2021product,tacchella2021relatedness}, the similarity between products was used to predict countries' future product exportation competitiveness. We can apply the general similarity measure defined in literature~\cite{teece1994understanding} to our MA-technology networks as: 
\begin{equation}
    B^y_{tt'}= \frac{1}{N_1}\sum_a \frac{M^y_{at}M^y_{at'}}{N_2},
\label{eq:sim1}
\end{equation}
in the case of technology similarity (between items), or
\begin{equation}
    B^y_{aa'}= \frac{1}{N_1}\sum_t \frac{M^y_{at}M^y_{a't}}{N_2},
\label{eq:sim2}
\end{equation}
in the case of similarity of MAs (between users). Here $N_1$ and $N_2$ are two parameters through which it is possible to define several types of similarity.\\
The simplest type is called \textit{co-occurrence}~\cite{teece1994understanding}, and it is defined putting $N_1 = N_2 = 1$. Given two nodes of the same layer, this measure counts how many common neighbour nodes they have in the other layer. In our case, we measure how many MAs do the technology $t$ and $t'$ in the same year or how many technologies are done by both MAs, $a$ and $a'$, in the same year. However, different similarity measures can be found in the literature based on the value given to $N_1$ and $N_2$. We define by $d_a = \sum_t M_{a,t}$ the diversification of the MA $a$, i.e., the number of technologies done by $a$, and by $u_t = \sum_a M_{a,t}$ the ubiquity of technology $t$, i.e., the number of MAs active in that technology sector. Among the broadest similarity measures used are:
\begin{itemize}
 \item Technology Space (TS). This similarity is based on the Product Space of~\cite{hidalgo2007product} and it has $N_1 = \max(u_t, u_{t'})$ and $N_2 = 1$ (or $N_1 = \max(d_a, d_{a'})$ and $N_2 = 1$ in the MA layer). Using this type of normalisation, one gives a lower connection weight to those technologies done by many MAs;
\item Resource Allocation (RA)~\cite{zhou2009predicting}. This similarity is obtained with $N_1 = 1$ and $N_2 = d_a$ ($N_1 = 1$ and $N_2 = u_t$ for MA layer). It is used to modulate the contributions of common neighbours with high degrees. If a MA has high diversification, RA will penalise the link between its technologies, given the triviality of their link. If the MA makes all the technologies, it is a given that each technology is linked with all the others.;
\item Taxonomy (TAX)~\cite{zaccaria2014taxonomy}. For this similarity $N_1 = \max(u_t, u_{t'})$ and $N_2 = d_a$ ($N_1 = \max(d_a, d_{a'})$ and $N_2 = u_t$ for the MA layer). The Technology Space gives a higher similarity score to technology with a low ubiquity (i.e., technology done by a few MAs) and, consequently, bias towards them. However, the idea is that these complex technologies are done by MAs (a few numbers) that do approximately all the others. Consequently, it is impossible to justify a city's path from non-complex technologies to complex ones. Normalising also for the diversification, we avoid this problem as we penalise low ubiquity scores and complex technologies are weighted more.
\end{itemize}
Following Hidalgo et al. \cite{hidalgo2007product}, we define the quantities:
\begin{equation}
    \omega^{tec}_{at} = \frac{\sum_{t'} M_{at'}B_{tt'}}{\sum_{t'} B_{tt'}}\ \ \ \ \ \ \ \ \ \ \
    \omega^{MA}_{at} = \frac{\sum_{a'} M_{a't}B_{aa'}}{\sum_{a'} B_{aa'}}.
\label{eq:pred}
\end{equation}
$\omega^{tec}_{at}$ measures how much the technologies done by the MA $a$ are similar to the technology $t$. $\omega^{tec}_{at}$ is thus high if MA $a$ develops technologies close to the technology $t$
$\omega^{MA}_{at}$, instead, measures how much a given technology $t$ is spread among MAs similar to the MA $a$. $\omega^{MA}_{at}$ is thus high if technology $t$ is spread among MAs surrounding MA $A$).\\
Given these definitions, we can use $\omega^{tec}_{at}$ ($\omega^{MA}_{at}$) as a prediction score: higher is $\omega^{tec}_{at}$ ($\omega^{MA}_{at}$), the higher the probability that an MA $a$ will start developing the technology $t$.

\subsection*{Testing Procedure}
Given a matrix $\textbf{M}^y$, one of our purposes is to predict the same matrix $\delta$ years later, $\textbf{M}^{y+\delta}$. The basic idea is that higher values in $\omega^{tec}_{at}$ or $\omega^{MA}_{at}$ will correspond to new technologies, i.e., more 1s, in $\textbf{M}^{y+\delta}$. To this end, we have to keep into account two elements.
\begin{itemize}
\item Class Imbalance. We are treating our problem as a classification one, i.e., we want to predict if a MA will do or not a given technology. Class labels, in our case, are 0s and 1s, but the number of elements equal to 1 is approximately only 5\%. To treat this unbalance correctly, we adopted the Area Under the Precision-Recall Curve~\cite{saito2015precision}.
\item Autocorrelation. With the term \textit{autocorrelation}, we mean that if a MA does or does not do a given technology in a specific year, with a high probability it will continue his current behaviour in the future. To avoid this problem, the evaluation is performed only for activations events, i.e., events in which the technology is not done in the year $y$ and it is done at year $y + \delta$. This strategy allows the healing of autocorrelation problems. Furthermore, it helps us study the diffusion of the technological process. We are more interested, in fact, in understanding where a new technology will be triggered, rather than knowing which ones will not.
\end{itemize}

%%%%%%%%%%%%%%%%%%%%%%%%%%%%%%%%%%%%%%%%%%
\section{Results}\label{sec:results}
\subsection{Predictions}

\subsubsection*{Geographic proximity and country diffusion}
\begin{figure}[h!]
    \centering
    \includegraphics[scale=.55]{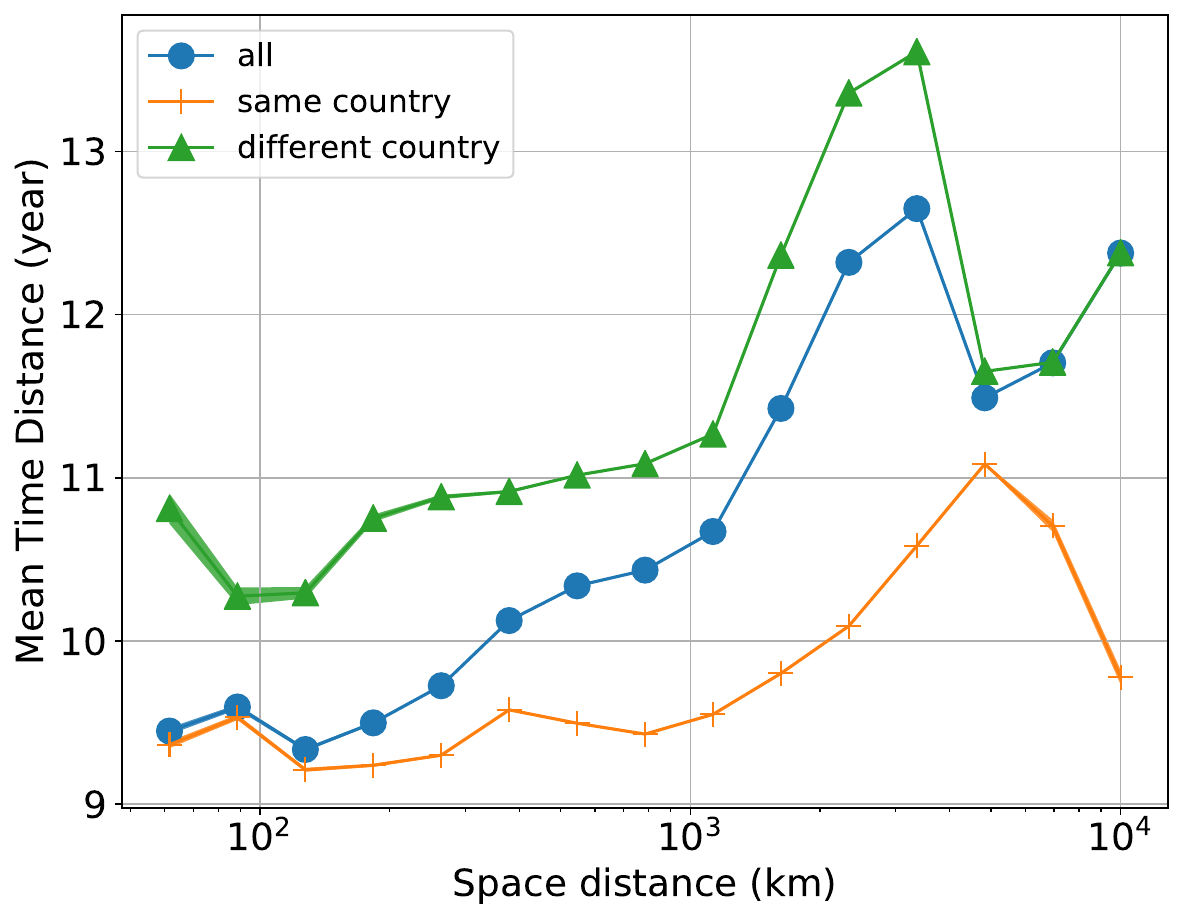}
\caption{Mean Time Distance (see the text for the definition) aggregated on different spacial distance ranges belonging. The error for each beam is determined by calculating the mean standard deviation. Due to the significant number of points per beam, the error is often not visually discernible in most plots. The blue curve corresponds to the aggregation of all MAs. We observe the overall increase in the Mean Time Distance, signalling an important role of geographical distances. Second, we split the set of MAs into two subsets of pairs of MAs belonging (orange curve) or not (green curve) to the same country. The second important observation is that belonging to the same country greatly reduces diffusion times.}
\label{fig:mTmS}
\end{figure}

We analyse technology code diffusion timing to study the role of physical and political geography in innovation dynamics. Consider the MA where a specific technology code $t$ first appears. We define the \textit{Mean Time Distance} as the average time distance between the first appearance of $t$ and its other first appearances in other MAs. After averaging over all technologies, we aggregate this mean on different spatial distance ranges to analyse the relationship with physical geography. On the other hand, to consider political geography, we calculate the average on the subsets of MAs belonging or not to the same country. In Fig.~\ref{fig:mTmS}, we report our analysis on the Mean Time Distance. 

Two important observations are in order. First, for the overall set of MAs, the Mean Time Distance increases on average with the geographical distance, signalling an important role of geography in the diffusion of innovation. Second, the Mean Time Distance is always shorter for the subset of MAs belonging to the same country, and it does not show a strong dependence from the spacial distance until the scale $10^3$ Km. After this scale, we see how a dependency from the spacial distance is stronger but more fluctuating (growing and then decreasing). This evidence is probably due to the distribution of MAs' distances, which are affected by seas and oceans. In fact, until the scale $10^3$ Km, the distribution of distances (presented in Supplementary Information) follows a power law with exponent $\sim 2$, corresponding to an isotropic distribution in two dimensions. After that scale, the seas and oceans break the isotropy assumption, making the distribution less predictable and ultimately affecting Mean Time Distance. But also in this range, the MAs couples from the same country show a way lower Mean Time distance. Therefore, we can consider political geography as predominant over physical geography in the dynamics of innovation.

\subsubsection*{Role of countries: an improved model}

In works concerning similarity and forecast on bipartite networks, it's common to compute the prediction using the links between the items layer (technology codes, in our case), i.e., using $\omega^{tec}_{at}$. However, mathematically, we have seen that it is possible to calculate a similarity between the nodes of both layers, i.e., also considering $\omega^{MA}_{at}$. In the work of Albora et al.~\cite{albora2022sapling}, the authors show how a mean between the two scores can outperform the standard method. They also propose a linear combination of item-based and user-based estimations, showing how this method outperforms the others. In our case, to get the prediction, we utilised this last method computing a linear combination of technology and MA densities instead:
\begin{equation}
    S^{y+\delta}_{at} = \alpha\omega^{tec}_{at} + \beta\omega^{MA}_{at}.
\end{equation}
where $S^{y+\delta}_{at}$ is the forecast for the year $y + \delta$. If we consider MAs with no patent in year $y$, regardless of the similarities used, the predictions obtained from $\omega^{tec}_{at}$ and  $\omega^{MA}_{at}$ will always be zero by construction. This outcome is due to the presence, in the rows of $\textbf{M}$ matrices related to those MAs, of only $0$s. Given the relevance of belonging to a country unveiled through our previous results, we included that information to predict when a given MA will start patenting a specific technology for the first time. To this end, we define:
\begin{equation}
    \omega^{C}_{at} = \sum_{a'}{M_{a't}^{y}\frac{C_{aa'}}{\sum_{a}{C_{aa}}}},
\end{equation}
where $C_{aa'} = 1$ if $a$ and $a'$ belong to the same country, $0$ otherwise and $\sum_a C_{aa}$ is the number of MAs in the same country as $a$, inserted to avoid size effects. $\omega^{C}_{at}$ represents the average values of technologies done by the MAs of a specific country. As explained in the Method section, the higher the value of $\omega^{C}_{at}$ is, the higher the probability that $M_{at}^{y+\delta} = 1$.\\
Our prediction  model is thus a linear combination of the three previous contributions: technology similarity, MA similarity and information on belonging to the same country:
\begin{equation}
    S^{y+\delta}_{at} = \alpha\omega^{tec}_{at} + \beta\omega^{MA}_{at} + (1-\alpha-\beta)\omega^{C}_{at}.
    \label{eq:tot_pred}
\end{equation}
Also in this case, the higher the value of $S^{y+\delta}_{at}$, the higher the probability to have $M_{at}^{y+\delta} = 1$. Because of the Autocorrelation problem explained in the Method section, we decided to evaluate our predictions on the so-called \textit{activation} elements, i.e., the matrix elements $M_{at}^{y} = 0$ and that in $y+\delta$ could become 1.\\

\begin{figure*}[h!]
\centering
\includegraphics[width=.75\textwidth]{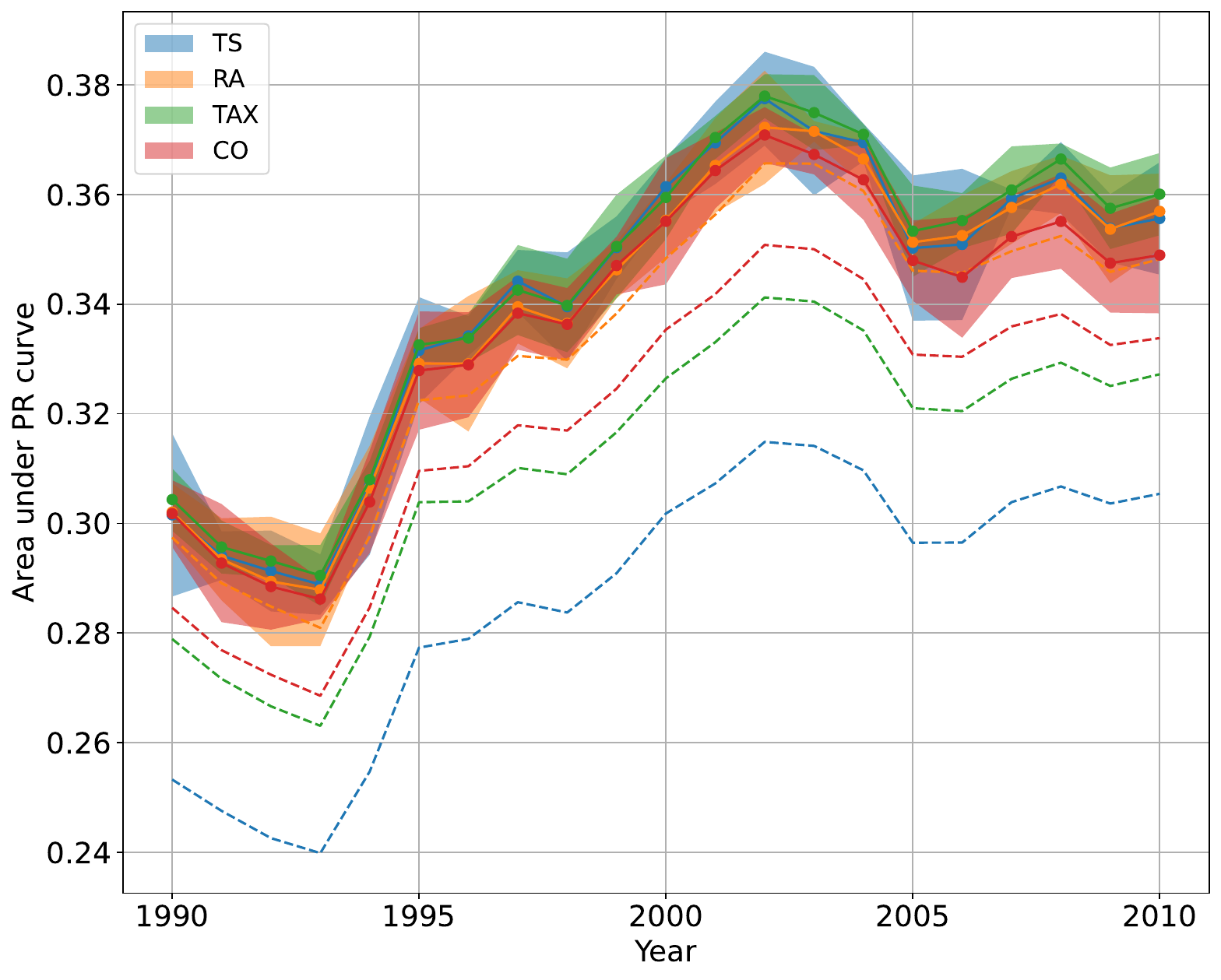}
%\subfloat[]%[\emph{}] {\includegraphics[width=.45\textwidth]{heat_map_alpha_beta_correction_newdata_TAX.pdf}} \\
%\subfloat[]%[\emph{}] {\includegraphics[width=.45\textwidth]{importance_country_correction_newdata.pdf}}
\caption{\textbf{Performances of predictions models.} Continue curves represent the prediction scores of our improved model (Eq.~\ref{eq:tot_pred}) for the four similarity metrics defined in the text, TS, RA, TAX and CO. For comparison, dotted curves report the same prediction scores of the classical model based on the item-item similarity $\omega^{tec}_{at}$. Our improved model outperforms the classic approaches. Error ranges are obtained using a 5-fold cross-validation to select the best parameter values $\Bar{\alpha}$ and $\Bar{\beta}$ out-of-samples.}
\label{fig:model}
\end{figure*}

In Fig. \ref{fig:model}, we compare the prediction for $\delta = 10$ of the four metrics of similarity defined above. We also compare our model (continue curves) and classic models, i.e., models using the items-items similarity $\omega^{tec}_{at}$ (dotted lines). 
%In our models, error ranges are obtained using 5-fold cross-validation to select the best parameter values $\Bar{\alpha}$ and $\Bar{\beta}$ out-of-samples. In the end, we compute the mean and the standard deviation. 
We can see how our model curves outperform all the dotted ones. In Supplementary Information, we also report the analysis done by using $\delta = 1$ and $\delta = 5$.\\
If we consider MA with no technologies in $y$, both $\omega^{tec}_{at}$ and $\omega^{MA}_{at}$ are $0$ by definition. In this case, the predictions of our models are only due to $\omega^{C}_{at}$, which represents the influence of countries.\\
In this specific case, we compared our results (Model) against a null model (Rand) and a model based on the spatial distance (Dist) to validate our findings. The null model prediction for each MA is a redistribution of the predicted technologies in the whole vector of the technological codes. If, for a given MA, we predict $(0,0,1,0)$, the null model would predict $(0.25,0.25,0.25,0.25)$. On the other hand, the spatial distance model uses geodetic distances between MA as similarities. In Tab. \ref{tab:geog}, we compare, for different values of $\delta$, the models' performances on technological debuts of MAs by summing the areas under the curves for all years. Our model, informed on country membership, is the most successful in estimating future technologies made by an MA with a null technology portfolio. %In Supplementary Information, we present the prediction curves for different values of $\delta$ for MAs with 0s technologies.
\begin{table}[!ht]
    \centering
    \begin{tabular}{c|c|c|c}
        ~ & $\delta_1$ & $\delta_5$ & $\delta_{10}$ \\
        Model & 0.075 & 0.094 & 0.138 \\
        Dist & 0.052 & 0.079 & 0.102 \\
        Rand & 0.017 & 0.032 & 0.076 \\
    \end{tabular}
\caption{\textbf{Models comparison}. In the table, we compare, for different values of $\delta$, the values of the areas under the curves of the predictions made on the MAs with zero technologies using the information belonging to the same country, geographic distances, and the random case. Same-country membership is the information that most successfully gives us an estimate of future technologies made by an MA with 0 technology portfolio.}
\label{tab:geog}
\end{table}
\subsection{Model analysis}

\begin{figure*}[h!]
\centering
\subfloat[]%[\emph{}]
   {\includegraphics[width=.45\textwidth]{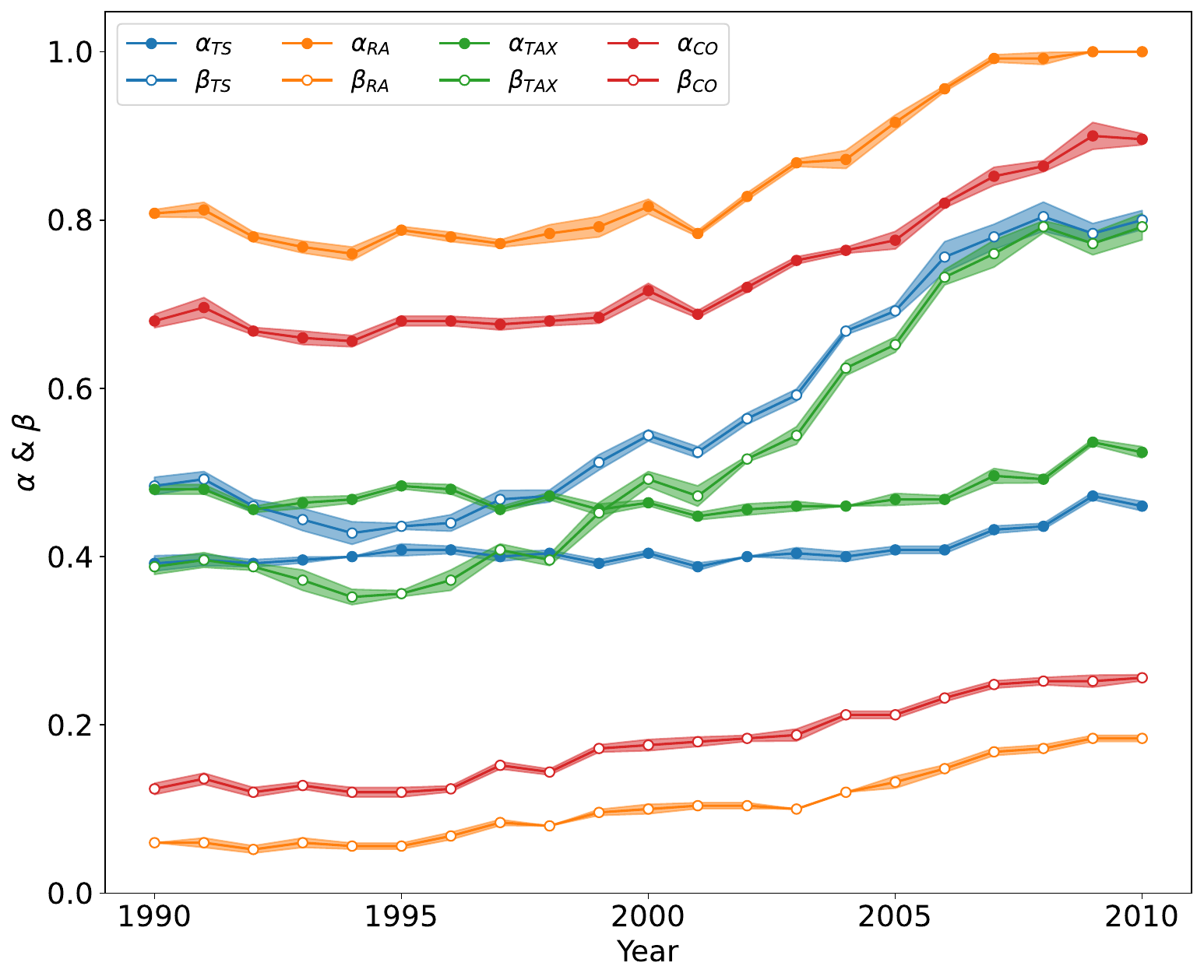}} \quad
\subfloat[]%[\emph{}]
   {\includegraphics[width=.45\textwidth]{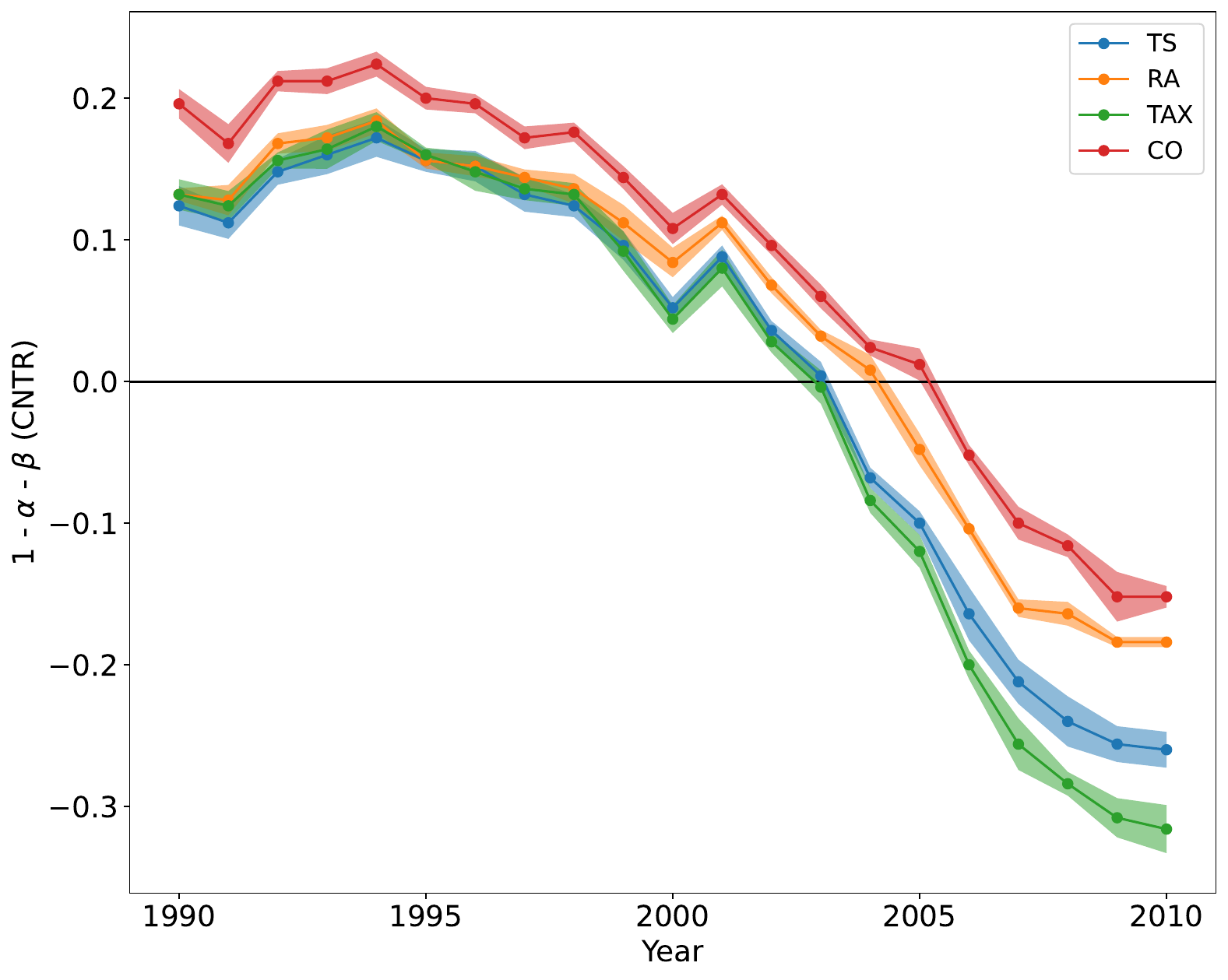}}
\caption{\textbf{Analysis of model optimal parameters with $\delta=10$}. \textbf{a}: Optimal $\alpha$ and $\beta$ over the years for different similarity metrics. We can see how both started to increase around 2000. \textbf{b}: The contribution of country information over the years, estimated as $1-\alpha - \beta$. We show how the contribution of country information is positive in the early years, but around the late 90s', this tends to decrease and even become negative.}
\label{fig:mod_an}
\end{figure*}

In this section, we analyse the behaviour of the best parameters $\alpha$ and $\beta$ over the years. For each metric, we show in Fig.~\ref{fig:mod_an}\textbf{a} the optimal values of $\alpha$ and $\beta$  over the years considering $\delta=10$. In Supplementary Information, we have reported the same analysis for $\delta=1$ and $\delta=5$. In this figure, we can see a common trend. Both $\alpha$ and $\beta$ tend to stay constant till the end of the 90s'. After that, their values tend to increase, as all four similarity metrics predicted. This analysis is confirmed by the descending behaviour, in Fig.~\ref{fig:mod_an}\textbf{b}, of the term $1-\alpha - \beta$, representing the importance of belonging to a country. These pieces of evidence suggest that political geography has been highly important for the diffusion of innovation till around two decades ago. After that, the evidence indicates that the overall ecosystem of MAs became more global and based more on similarities between technologies and MAs.
At the beginning of the data years, the country term $1-\alpha - \beta$ has a positive contribution, but around the end of the 90s', it tends to decrease and even becomes negative. We interpret this result as a change in the dynamics of innovation in countries where the similarity between technologies and MA starts to become more important than the country itself. This is likely because, instead of following national trends, many MAs could have begun to copy MAs in other countries.

\subsection*{The paths to innovation}

\begin{figure*}[h!]
    \centering
    \includegraphics[scale=.70]{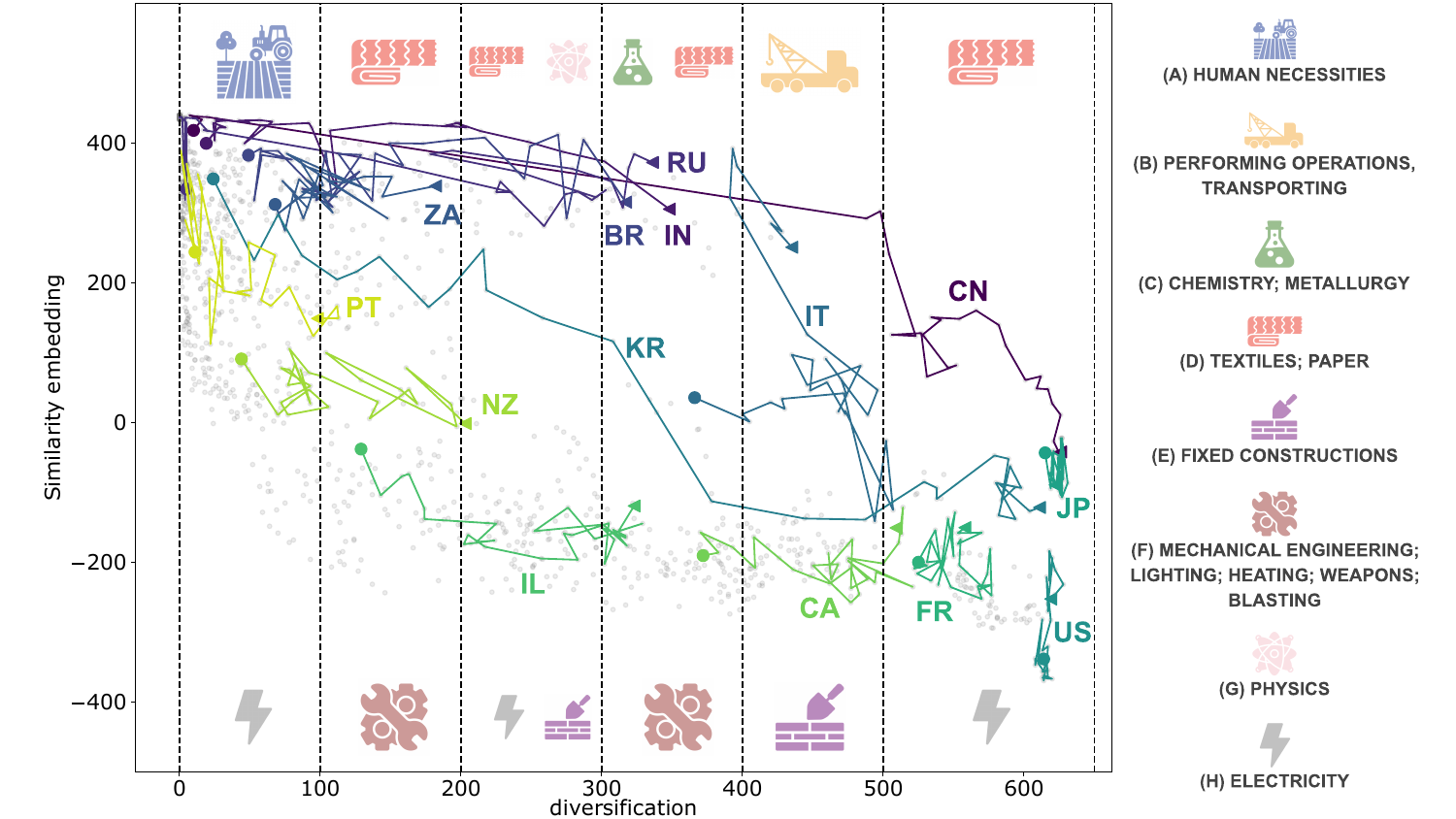}
\caption{\textbf{Country's 1D Similarity embedding vs. diversification.} Each point represents a country in a given year. For some countries, we plotted the trajectory over time. We can see how countries tend differently to reach a point of accumulation where the most developed countries are. In the lower part, we find the typical path of Western countries, and we report, for example, France, Canada, New Zealand and Israel. To highlight the technology difference between the ``upper'' and the ``lower'' paths, we divided the diversification into ranges of size 100 (except the last one). We focus on each range's highest and lowest 25th percentile, aggregate the technologies to the 1st digit, and identify the most distinctive of the two subsets. The relative icons are reported on the top and bottom of each diversification range. 
The ``upper'' part is dominated mainly by the BRICS, Russia, India, China and Brazil. In technology code terms, we can highlight the differences between the two extreme paths: the ``upper'' part dominates mostly in manufacturing technology as Textiles and Paper. The leftmost part, i.e., the least diverse, particularly dominates in technologies devoted to Human necessities. The ``lower'' part dominates in most sophisticated technologies such as Electricity, Fixed construction and Mechanical engineering.}
\label{fig:UMAP_1D_CNTR}
\end{figure*}

\begin{figure*}[h!]
    \centering
    \includegraphics[scale=.70]{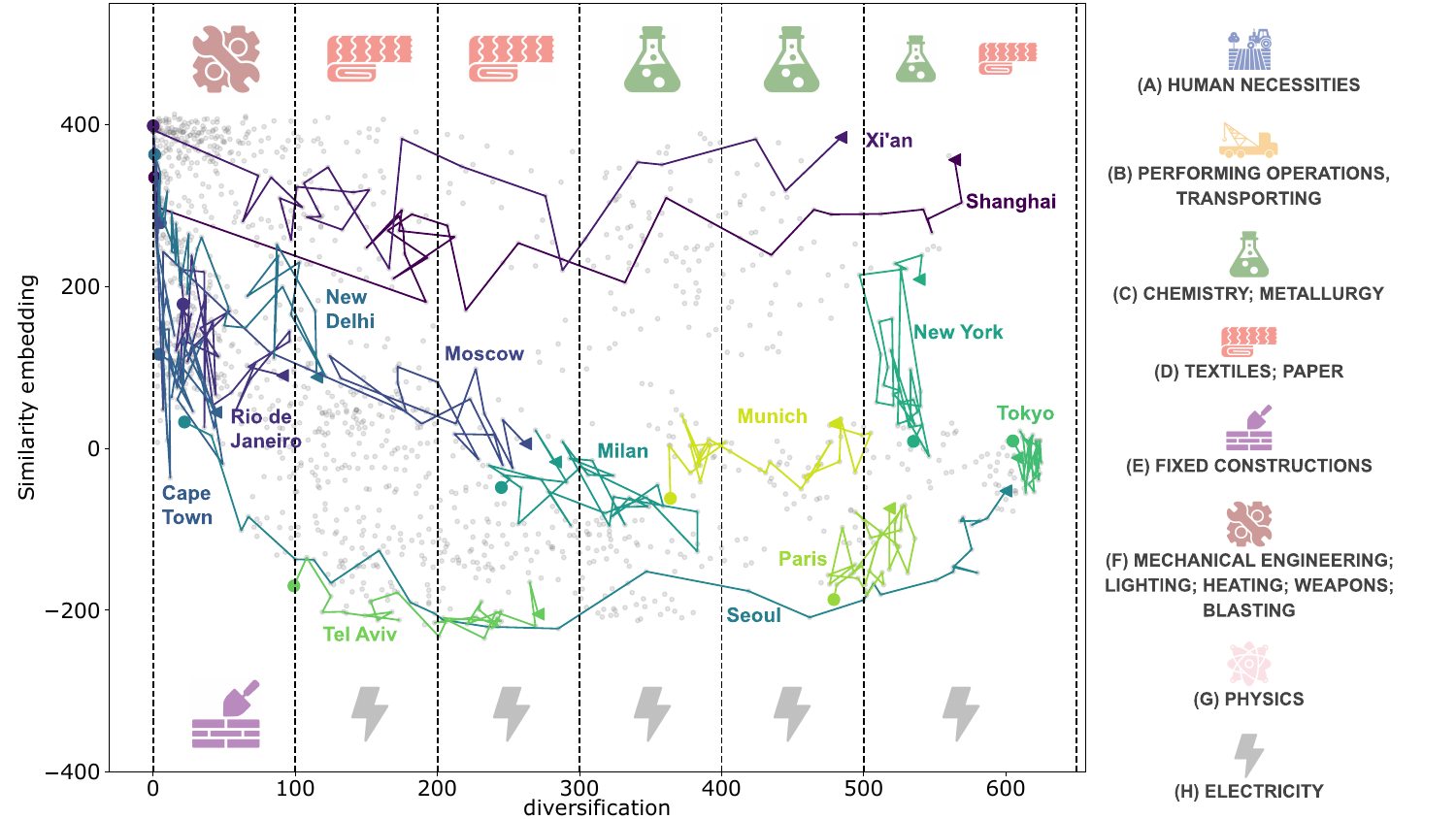}
\caption{\textbf{MA's 1D Similarity embedding vs. diversification.} Each point represents a MA in a given year. To highlight the technology difference between each diversification range's ``upper'' and the ``lower'' paths, we follow the same procedure of Fig.~\ref{fig:UMAP_1D_CNTR}. The technology differences show that the lower path dominates in Electricity technologies, while the upper path dominates in Chemistry, Textiles and Paper technologies. We see how some MAs tend to chase others (Seoul vs Tokyo, and Moscow vs Milan), though, unlike the countries' case, no single accumulation point emerges.}
\label{fig:UMAP_1D_MA}
\end{figure*}

In this last section, we focus on innovation paths, i.e., the paths followed by countries and metropolitan areas towards technological innovation. Though diversification is a good proxy for progress to innovation, we need another metric to represent similarities between the countries' development strategies. We define, in particular, a metric that quantifies how competitive a country $c$ is in a specific technology code $t$ in year $y$ relative to other countries, based on the number of MAs in $c$ that patent with that technology code. Similarly, we can quantify how competitive a MA $a$ is compared to other MAs. For each country, we define the following:
\begin{equation}
    G_{ct}^y= \frac{C_{ct}/C_c}{C_{wt}/C_w},
    \label{eq:stracca1}
\end{equation}
$C_{ct}$ counts how many MAs in the country $c$ do the technology $t$, and $C_c$ is the number of MAs in the country $c$. $C_{wt}$ counts how many MAs are in the entire database patent with the technology code $t$, and $C_w$ is the total number of MAs. Therefore, $G_{ct}^y$ measures the fraction of MAs in $c$ that do the technology $t$ compared to the entire word for the year $y$. We define with $\Bar{G}_{c}^{y}$ the vector that represents the average of $G_{ct}^y$ over all technologies $t$, and it represents the competitive position of the country $c$ for the year $y$. Similarly, for each MA, we  define the following:
\begin{equation}
    G_{at}^y= \frac{M_{a\in c,t}}{C_{ct}/C_c}.
    \label{eq:stracca2}
\end{equation}
and, similarly, $\Bar{G}_{a}^{y}$ is the average of $G_{at}^y$ over all technologies $t$ and it represents the competitive position of MA $a$ for the year $y$. For every year, $G_{ct}^y$ and $G_{at}^y$ are vectors with 650 entries, corresponding to the total number of technologies. Using UMAP, we reduced the dimensionality to one and defined the similarity embedding. We found that this embedding is strongly anti-correlated with the modules of $G_{at}$ and $G_{ct}$ (see the Supplementary Information for further information). This evidence implies that the lower the similarity embedding, the higher the competitiveness of countries or MAs. We can thus use the similarity embedding as a reverse measure of competitiveness and plot the time evolution of each country and each MA in a two-dimensional scatter plot determined by the two quantities: similarity embedding (a reverse proxy for competitiveness) and diversification. We report the results in Fig.~\ref{fig:UMAP_1D_CNTR} for countries and Fig.~\ref{fig:UMAP_1D_MA} for metropolitan areas. Each point on the two plots is a pair country/year and MA/year. \\

We have highlighted the paths over time, followed by a selection of countries and MAs. Two typical patterns emerge that we denote as the "upper" path and the "lower" path. This pattern is particularly evident for countries. A country or MA that moves from left to right increases its diversification but not the competitiveness in the technologies that it does. Instead, movements from the upper part to the bottom are associated with growth in terms of competitiveness, keeping fixed diversification. The main difference between the two typical paths is the order of these movements. In the ``upper'' path, we first observe an increasing diversification and then an increase in competitiveness. In the ``lower'' path, the opposite occurs: first, an increase in competitiveness followed by a diversification increase. We coloured with different shades of the same colour the evolution of some countries belonging to the two typical paths.\\
Finally, to highlight the technology difference between the ``upper'' and the ``lower'' paths of both figures, we divided the diversification into ranges of size 100 (except the last one). For each range, we focus on the highest and lowest 25th percentile and aggregate the technologies to the 1st digit, representing the general technological category. We compare the technological categories present in the two sets to highlight the most distinctive ones, i.e., those with the greatest difference in rank based on their frequency in the subset. For instance, if a technological category $X$ is the most common in the top 25\% set and the least common in the bottom 25\% set, $X$ will be considered as distinctive of the top set while, if it had been the most common in both sets, it would not have been considered distinctive. See Supplementary Information for more details.\\

In Fig.~\ref{fig:UMAP_1D_MA}, we show the results for MAs. Unlike countries, we do not observe a point of accumulation between MAs. We observe how some MAs get closer to others, such as Moscow to Milan, Seoul to Tokyo or Shanghai to New York. From a technological point of view, results are consistent with countries. The upper part is dominated by manufacturing technologies, while at the bottom one is more evident dominance of Electricity technologies.

%%%%%%%%%%%%%%%%%%%%%%%%%%%%%%%%%%%%%%%%%%
\section{Discussion}\label{sec:discussion}
This study provides valuable insights into technology diffusion among MAs worldwide and how geography impacts this process. Comparing geographic proximity, we find that belonging to a country is relevant in determining the likelihood of technology diffusion between metropolitan areas. Results indicate that, at equal geographical distances, technology diffusion occurs more readily across metropolitan areas belonging to the same country.

We develop a predictive model for future technology production of MAs that considers similarities between technologies and metropolitan areas and adds the contribution related to belonging to the same country. This last term allows for predictions even for metropolitan areas with empty technology portfolios. Our model outperforms traditional algorithms, particularly when one focuses on the case of technological debuts, i.e.,  when a metropolitan area starts developing a technology for the first time.

The study of the forecasts and the models' parameters highlights the increasing importance of similarities between technologies and metropolitan areas as years pass. In particular, around the end of the 90s, belonging to a country lost its significance as a predictor of innovation paths in favour of the similarity among technologies and metropolitan areas. This finding suggests a change in the dynamics of innovation. To get a deeper insight into this phenomenology, we represented the temporal paths of MAs and countries in the technological space of innovations. This space comprises two dimensions, corresponding to technological competitiveness and the diversification of countries and metropolitan areas. We singled out two main paths, one followed by most Western countries and the other by the BRICS ones.

In Fig.~\ref{fig:UMAP_1D_CNTR} the presence of a main growth path (with the country as New Zealand, Israel, France, etc.) is evident. The upper part is instead dominated mainly by the BRICS, Russia, India, China, Brazil and South Africa. We can highlight the differences between the two paths in technology code terms: the upper part dominates mostly in manufacturing technology, as Textiles and Paper. The rightmost part, i.e., the least diverse, particularly dominates Human necessities technologies. The lower part dominates in most sophisticated technologies such as Electricity, Fixed construction and Mechanical engineering.

The model developed in this study can predict technology diffusion transparently and understandably, differently from other ``black box'' predictive models present in literature. These features allow for informed decision-making regarding investment and innovation. From this perspective, our scheme could be a valuable tool for policymakers to guide investment decisions and prioritise innovation areas.

On a scientific level, this study opens the door to future work and questions. First, starting from the model presented in this work, which is focused on activations, i.e., first occurrences of a given technology, one could generalise to predict also predict ``shutdowns'', i.e., when a technological category is not patented any more. Furthermore, model simulation can be used to build green and sustainable pathways and highlight them at the level of MAs, regions, countries or companies. Finally, the relationship between forecasts and macroeconomic variables such as GDP can be explored to improve our understanding of innovation and economic dynamics.

Before concluding, it is essential to understand the limitation of the model. The only use of patents as a proxy for innovation~\cite{hall2014choice} represents one of the constraints. Inventions do not represent all forms of knowledge production in the economy, nor do patents cover all generated knowledge~\cite{arts2013inventions}. Moreover, it has been argued that the disadvantage of using patents is that it is difficult to estimate their value~\cite{hall2005market}. Second, remote working and dispersed research teams can mitigate the concentration of innovation in urban areas~\cite{clancy2020case,delventhal2020spatial,gupta2022work,shearmur2012cities}, and future studies linked with this should take this phenomenon into account.

\bibliography{sample.bib}

\section{Data availability statement}
The data supporting this study's findings are available upon reasonable request from the authors.

\section{Acknowledgements}
The authors acknowledge the CREF project "Complessità in Economia".

%%%
\end{document}

\title{\huge{Supplementary Information}\\\Large The geography of innovation dynamics.}

\author{Matteo Straccamore, Pietro Gravino and Vittorio Loreto}

\date{}
\maketitle

\newpage

\tableofcontents
\newpage

\newpage

\section{Patent Data and Geolocation Methodology}

In this study, we utilize the comprehensive patent database provided by De Rassenfosse et al.~\cite{de2019geocoding}, which offers geolocation information for approximately 18.9 million patent applicants. Our objective is to construct a dataset consisting of early patent applications from across the globe, organized according to the geographical locations of the applicants. This dataset enables us to identify the origins of patented inventions and locate the centers of innovation. The geographic information provided by this dataset proves valuable for investigating the geography of innovation, understanding the spatial distribution of patented inventions, and informing policymakers interested in firm location decisions and the attraction of highly skilled workers.\\
The authors of the database perform geolocation by associating postal codes of applicant addresses with latitude and longitude coordinates, thereby linking the applicants to their respective countries, regions, and cities/metropolitan areas. To obtain the postal code information, the authors match the addresses from the patent applications in PATSTAT and other databases, which, in turn, provide the corresponding postal codes.\\
Here are some pertinent features of this database:

\begin{itemize}
\item The patents included in this dataset represent the first application for a given invention.
\item Some patents may have multiple geolocations due to collaborations among multiple applicants.
\item Each patent is associated with one or more IPC 4-digit technology codes. In cases where a patent is associated with multiple technology codes, we retain all of them. This approach helps preserve the granularity of the data, as the technology categories are already quite specific.
\item As the database focuses on the first application of a patent, when a patent is shared across multiple patent offices, the applicant's position remains consistent.
\item The geolocation information for all patents in the De Rassenfosse database is obtained from PATSTAT, WIPO, REGPAT, and the patent offices of Japan, China, Germany, France, and the United Kingdom.
\end{itemize}

\section{Distances distribution}
\begin{figure}[h!]
    \centering
    \includegraphics[scale=.60]{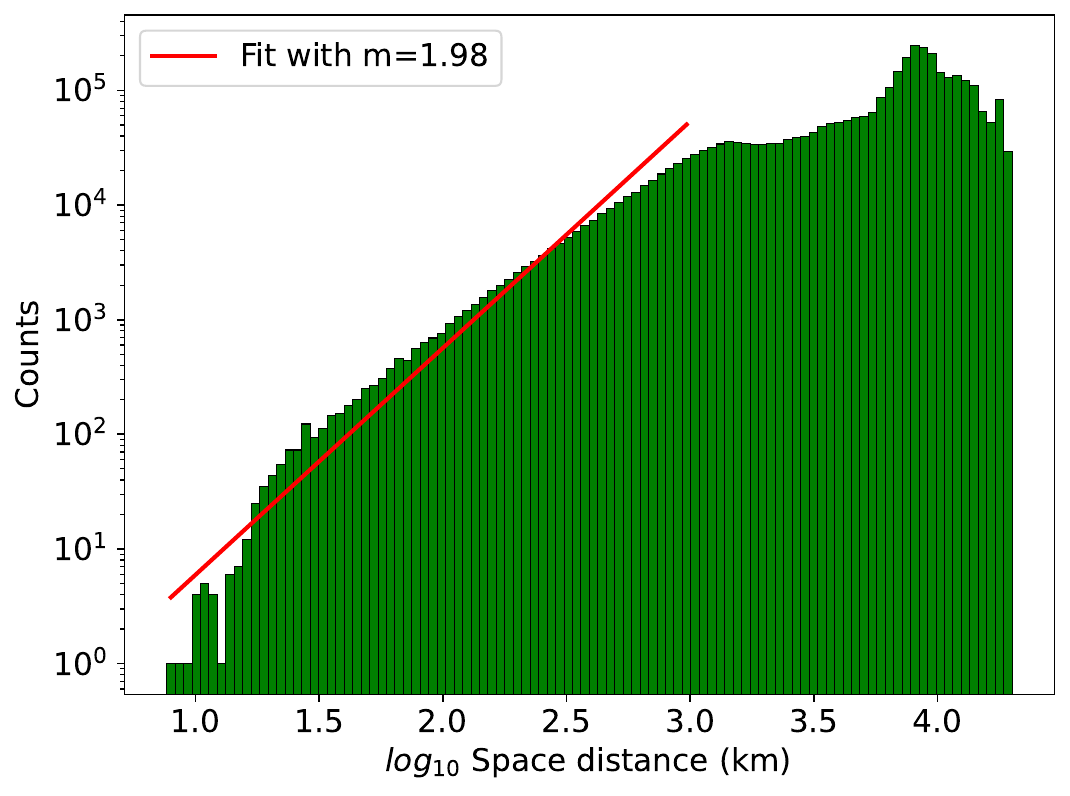}
\caption{\textbf{Distances distribution}. Distances distribution in log-log scale. Up to a distance of approximately $10^{3}$ km, the distribution exhibits a relatively weak reliance on spatial distance. However, beyond this threshold, the influence of spatial distance becomes more pronounced yet remains somewhat erratic, fluctuating with both increases and decreases. Within the range, the distance distribution follows a power law with an exponent of $\sim 2$, indicative of an isotropic distribution within two dimensions. The red line represents the fit of the distribution until $10^{3}$ km. The resulting angular coefficient is $1.98$.}
\label{fig:dist_distr}
\end{figure}
In order to gain a deeper understanding of the patterns depicted in Fig. 1 of the main text, we will now illustrate the distribution of distances in Fig \ref{fig:dist_distr}. Observing the graph, we can discern that up to a distance of approximately $10^{3}$ km, the distribution exhibits a relatively weak reliance on spatial distance. However, beyond this threshold, the influence of spatial distance becomes more pronounced yet remains somewhat erratic, fluctuating with both increases and decreases. Within the range, the distance distribution follows a power law with an exponent of $\sim 2$, indicative of an isotropic distribution within two dimensions. However, once this scale is surpassed, the presence of seas and oceans disrupts the assumption of isotropy, rendering the distribution less predictable. In the Figure, we plot also as a red line the fit of the distribution until $10^{3}$ km. The resulting angular coefficient is $1.98$.

\section{UMAP for visualization of innovation diffusion}
UMAP \cite{mcinnes2018umap} is a dimensionality reduction technique that is used to map high-dimensional data to a lower-dimensional space while preserving the structure of the data as much as possible. It uses a combination of techniques from algebraic topology and Riemannian geometry to construct a low-dimensional representation of the data, which can be useful for visualization and exploration. UMAP has been shown to be competitive with other dimensionality reduction techniques such as t-SNE \cite{van2008visualizing} and PCA \cite{mackiewicz1993principal}, while preserving more of the global structure of the data.\\
Here we describe the parameter that we compute to find the best representation:
\begin{itemize}
    \item \textbf{n\_components}: This parameter controls the number of dimensions in the low-dimensional representation. We set this parameter equal to 1.
    \item \textbf{metric}: This parameter determines the distance metric used for the data. We chose to use the Cosine metric;
    \item \textbf{n\_neighbors}: This parameter controls the number of nearest neighbours used to construct the low-dimensional representation. A larger value will preserve more of the local structure of the data, while a smaller value will result in a more global view of the data. The optimal value for this parameter will depend on the size and structure of the data. We put this value to 50;
    \item \textbf{min\_dist}: This parameter controls the minimum distance between points in the low-dimensional representation. A larger value will result in more spread-out points, while a smaller value will result in more densely packed points. This parameter can be used to control the level of clustering in the low-dimensional representation. We put this value equal to 0;
    \item \textbf{spread}: This parameter controls the degree of freedom of the low-dimensional representation. A larger value will result in a more evenly distributed representation, while a smaller value will result in a more tightly packed representation. This parameter can be used to control the level of clustering in the low-dimensional representation. We put this value to 1000;
\end{itemize}

\section{Predictions with different delay values.}
\begin{figure}
\centering
\subfloat[]%[\emph{}]
   {\includegraphics[width=.40\textwidth]{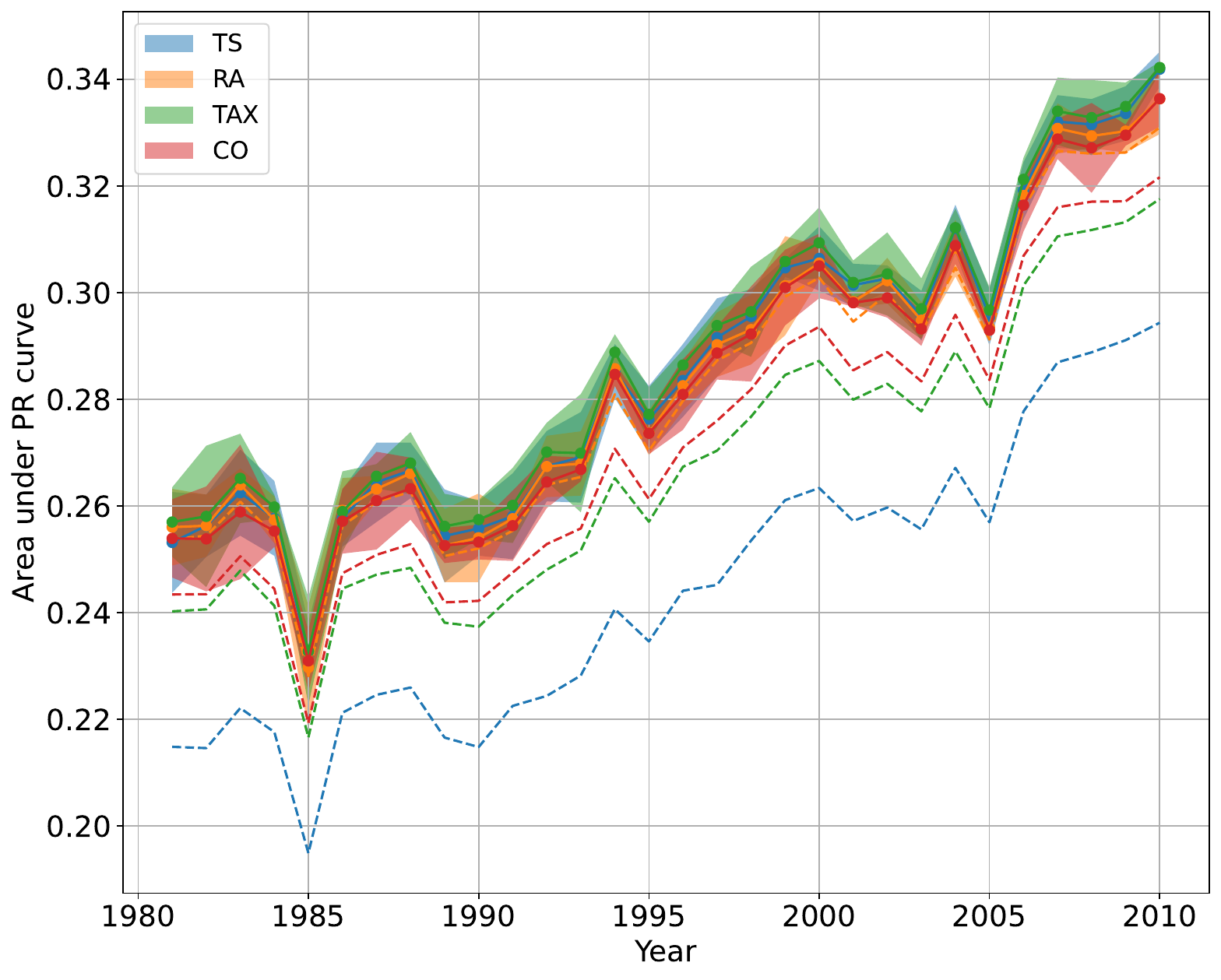}} \quad
\subfloat[]%[\emph{}]
   {\includegraphics[width=.40\textwidth]{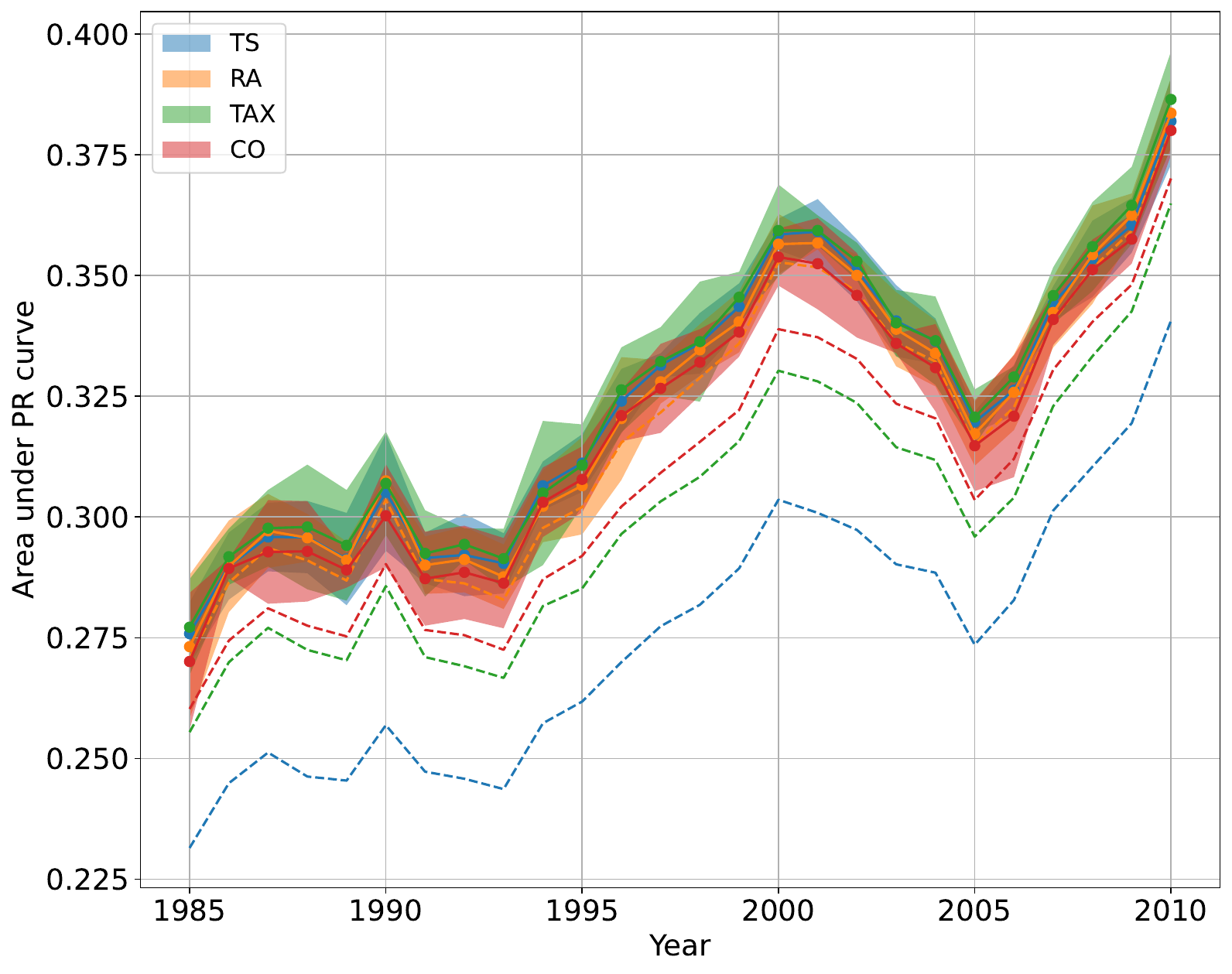}} \\
\subfloat[]%[\emph{}]
   {\includegraphics[width=.40\textwidth]{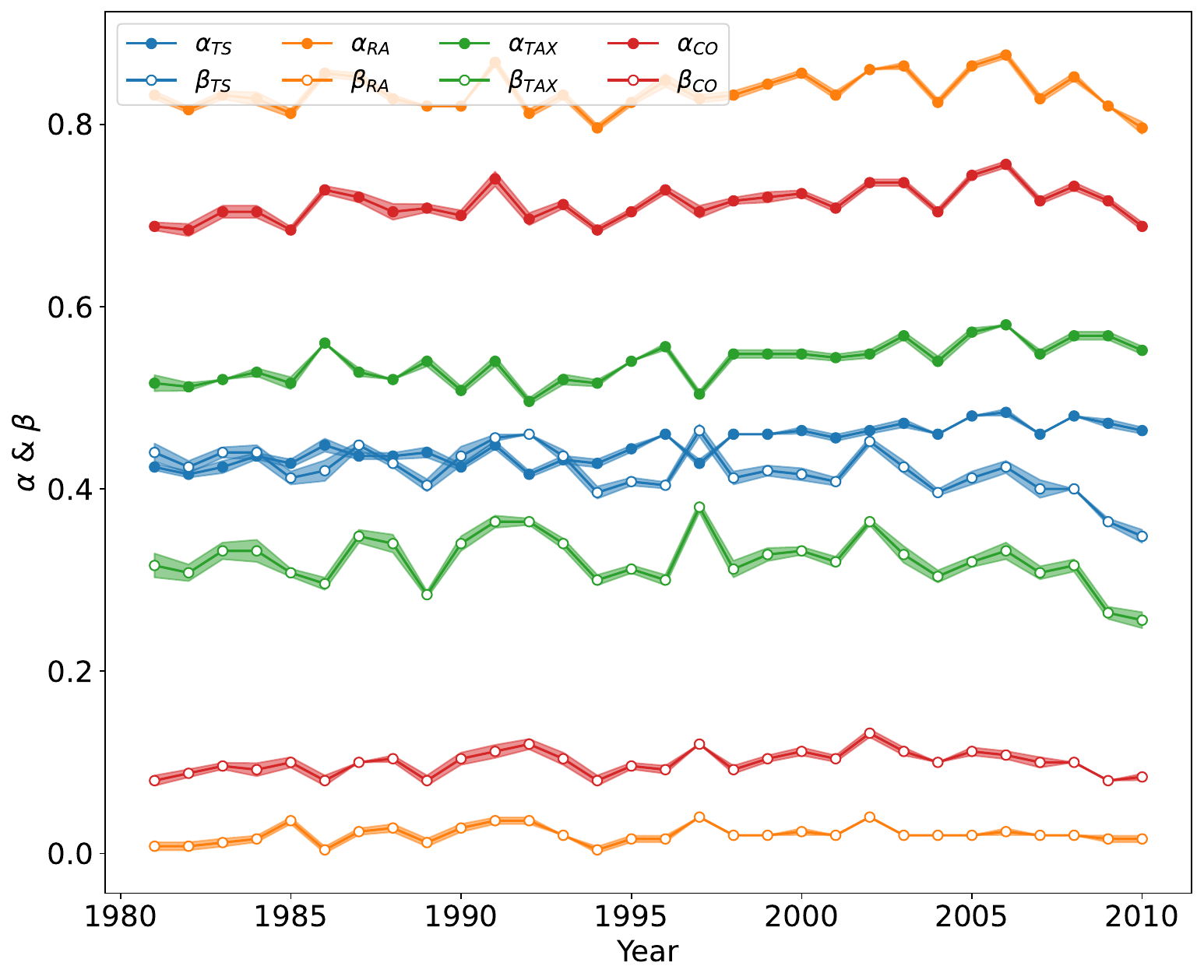}} \quad
\subfloat[]%[\emph{}]
   {\includegraphics[width=.40\textwidth]{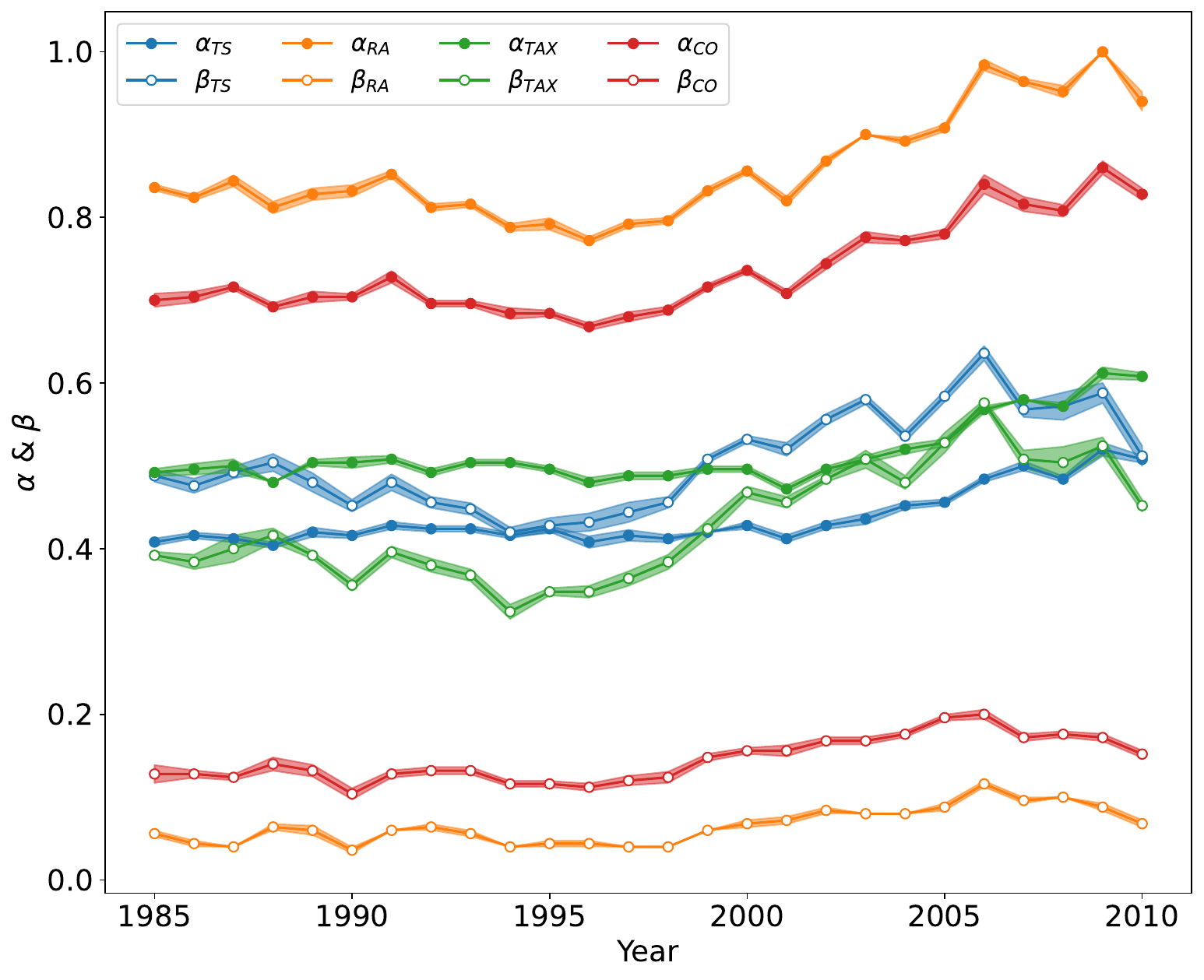}} \\
\subfloat[]%[\emph{}]
   {\includegraphics[width=.40\textwidth]{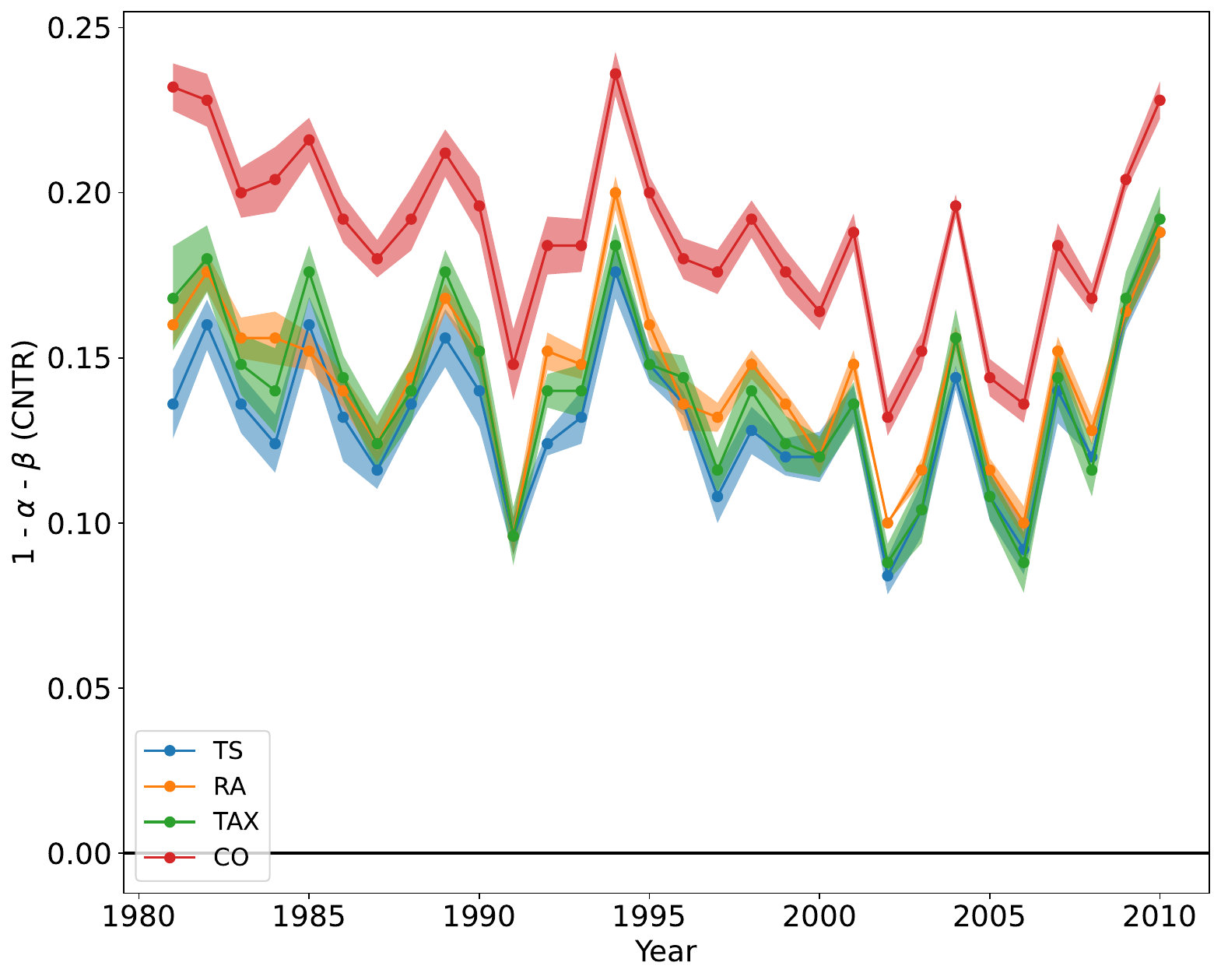}} \quad
\subfloat[]%[\emph{}]
   {\includegraphics[width=.40\textwidth]{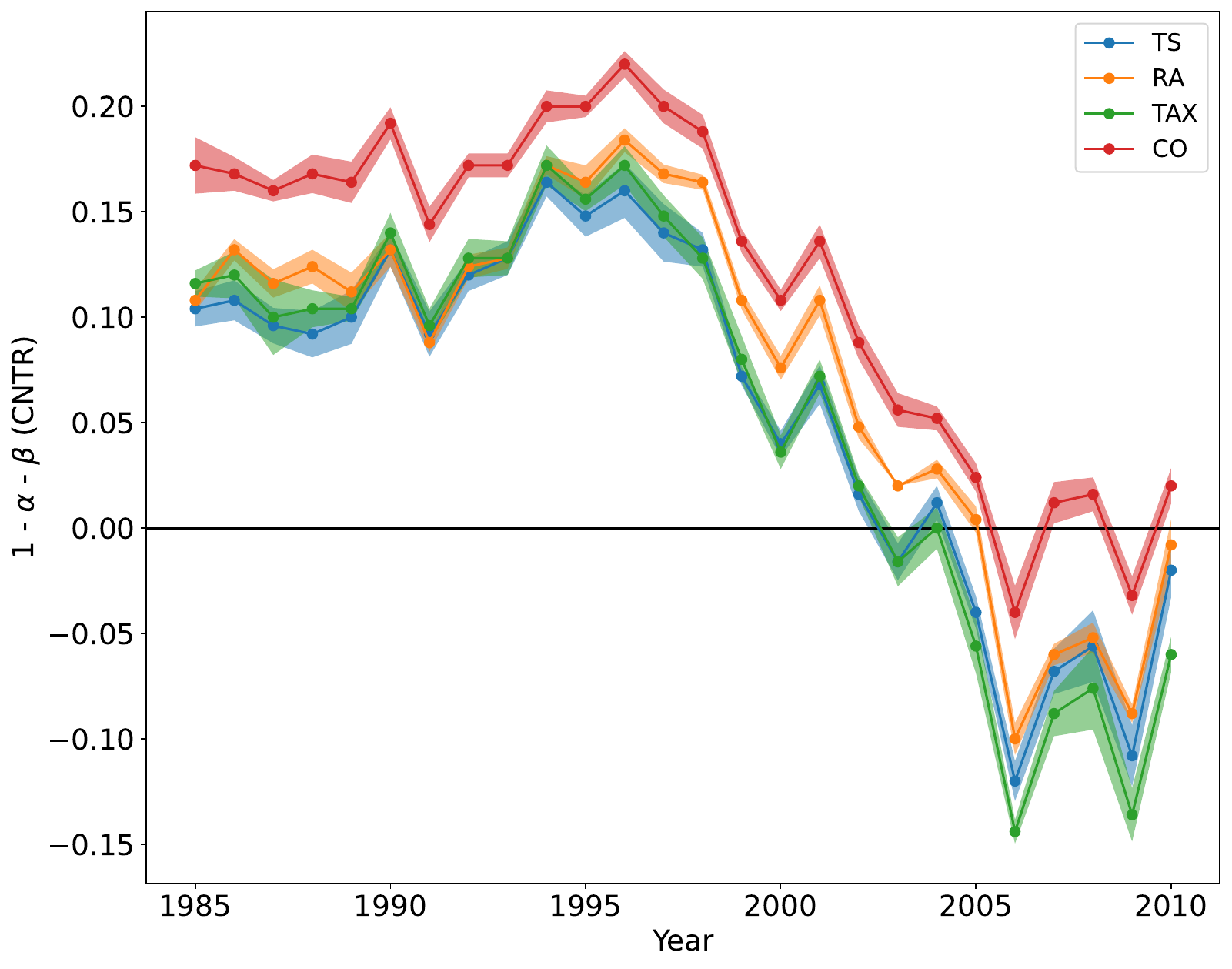}}
\caption{\textbf{$\delta = 1$ e $\delta = 5$ predictions with parameters analysis.} \textbf{(a-b)}: Predictions with $\delta = 1$ and $5$ respectively. In the case of $\delta=1$ we notice stability in parameter dependence, while already for $\delta=5$ we get consistent results with $\delta=10$ of the main text. The scores at $\delta = 1$ are still lower than at $\delta = 5$ and $\delta = 10$, and this is consistent with the result shown in Fig.1 of the main text in which we show that on average the time to the diffusion of technologies is about 10 years. \textbf{(c-d)}: Parameter trends of $\delta = 1$ and $5$ respectively. While in the figure of $\delta=5$, we notice similar but less pronounced behaviour than $\delta=10$, $\delta=1$ shows stable parameter values. \textbf{(e-f)}: Country importance of $\delta = 1$ and $5$ respectively. Also in this case, for $\delta=5$ we notice similar but less pronounced behaviour than $\delta=10$. Instead for $\delta = 1$, we show an almost stable behaviour.}
\label{fig:model_2}
\end{figure}
In this Section, we present the prediction results setting the $\delta$ equal to $1$ and $5$. In general, the results are coherent with each other. In Fig.\ref{fig:model_2} we show in each column respectively $\delta = 1$ and $5$; each figure row is referred to predictions, parameter trends and country importance. The $\delta = 5$ case presents a similar but less pronounced behaviour than $\delta=10$. Instead, $\delta=1$ appears more stable, especially in terms of parameters. However, we can see that the scores at $\delta = 1$ are still lower than at $\delta = 5$ and $\delta = 10$, and this is consistent with the result shown in Fig.1 of the main text in which we show that on average the time to the diffusion of technologies is about 10 years.

\section{Technology differences more information.}
\begin{figure}
\centering
\subfloat[]%[\emph{}]
   {\includegraphics[width=.38\textwidth]{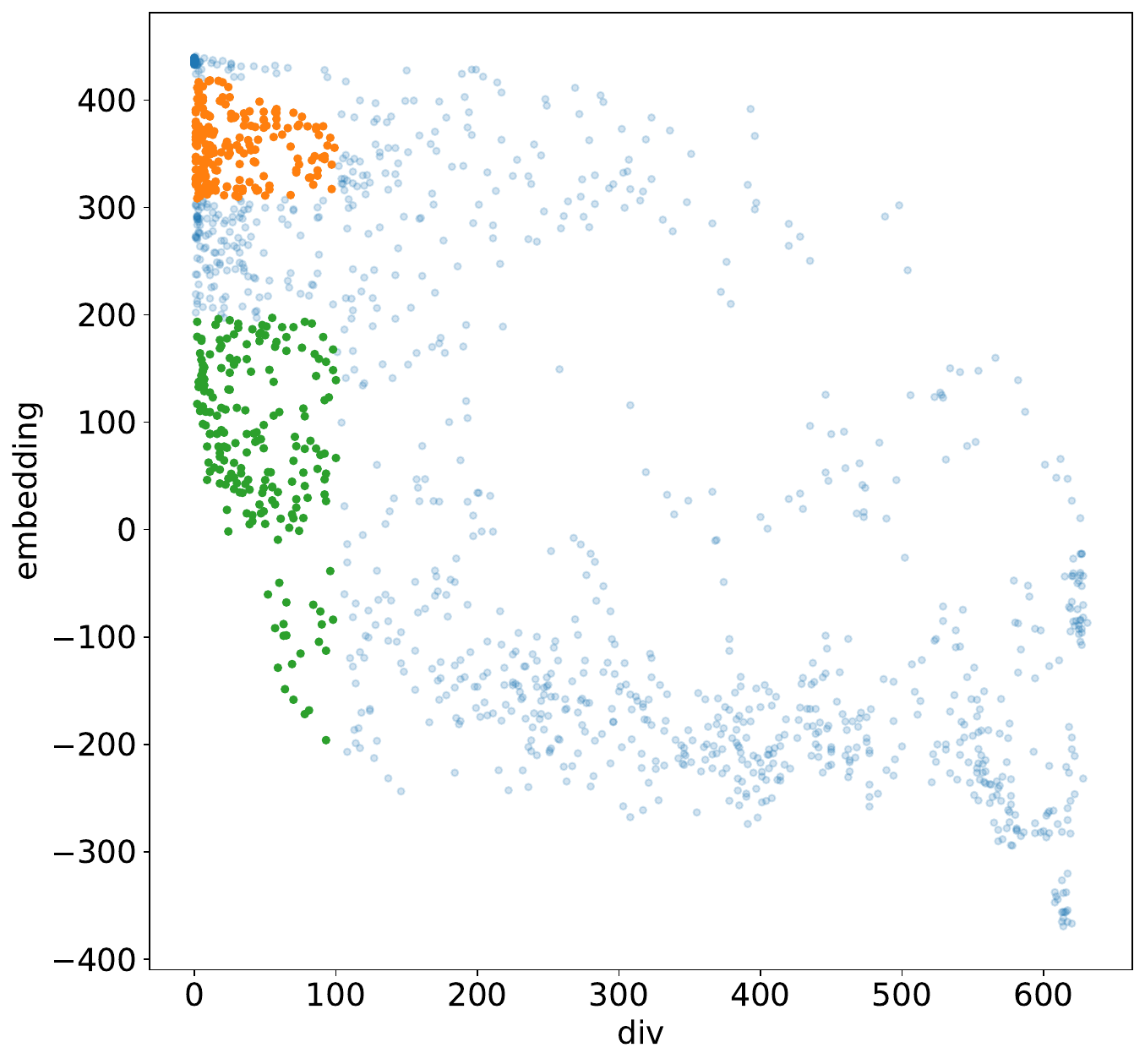}} \quad
\subfloat[]%[\emph{}]
   {\includegraphics[width=.38\textwidth]{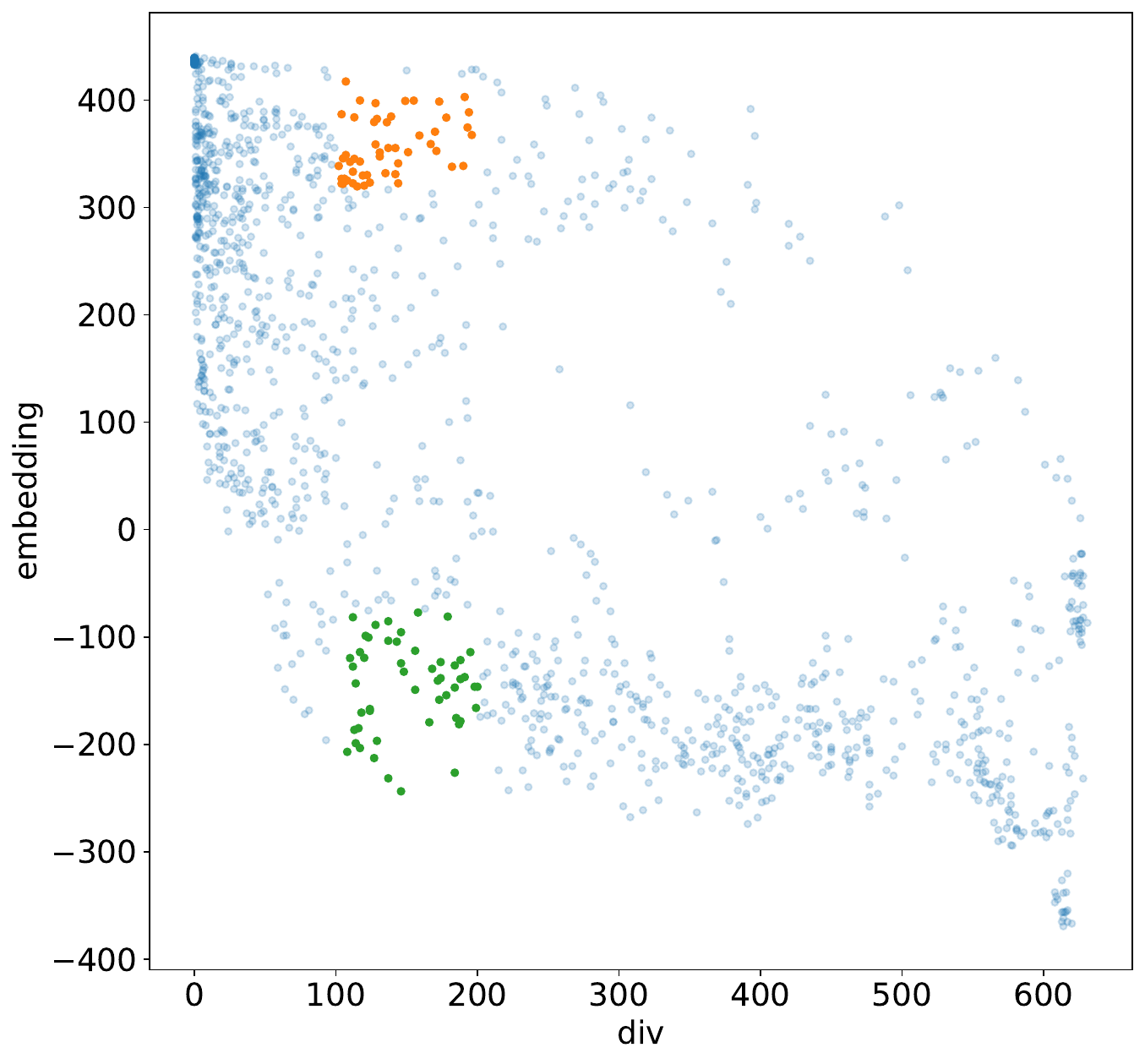}} \\
\subfloat[]%[\emph{}]
   {\includegraphics[width=.38\textwidth]{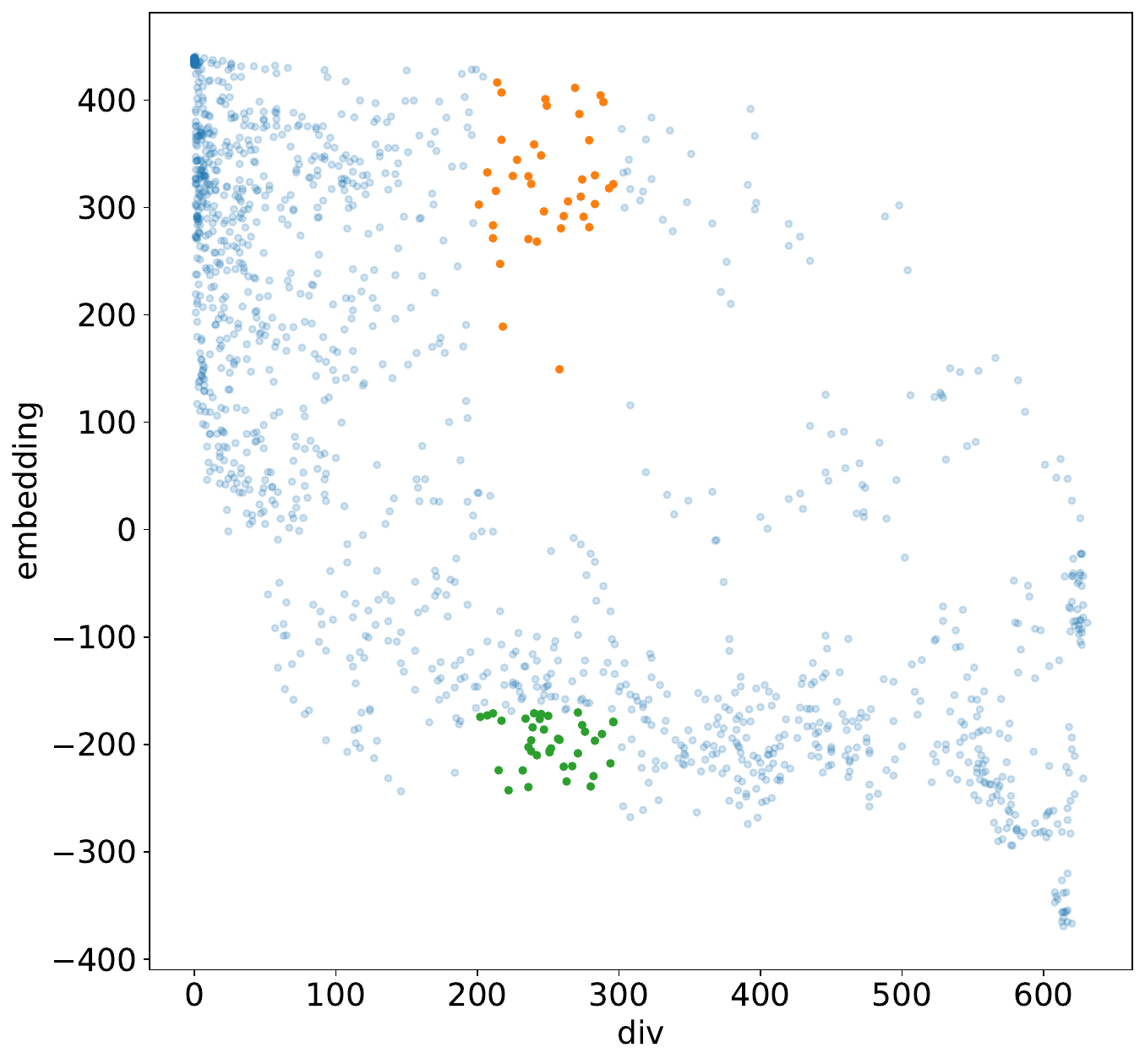}} \quad
\subfloat[]%[\emph{}]
   {\includegraphics[width=.38\textwidth]{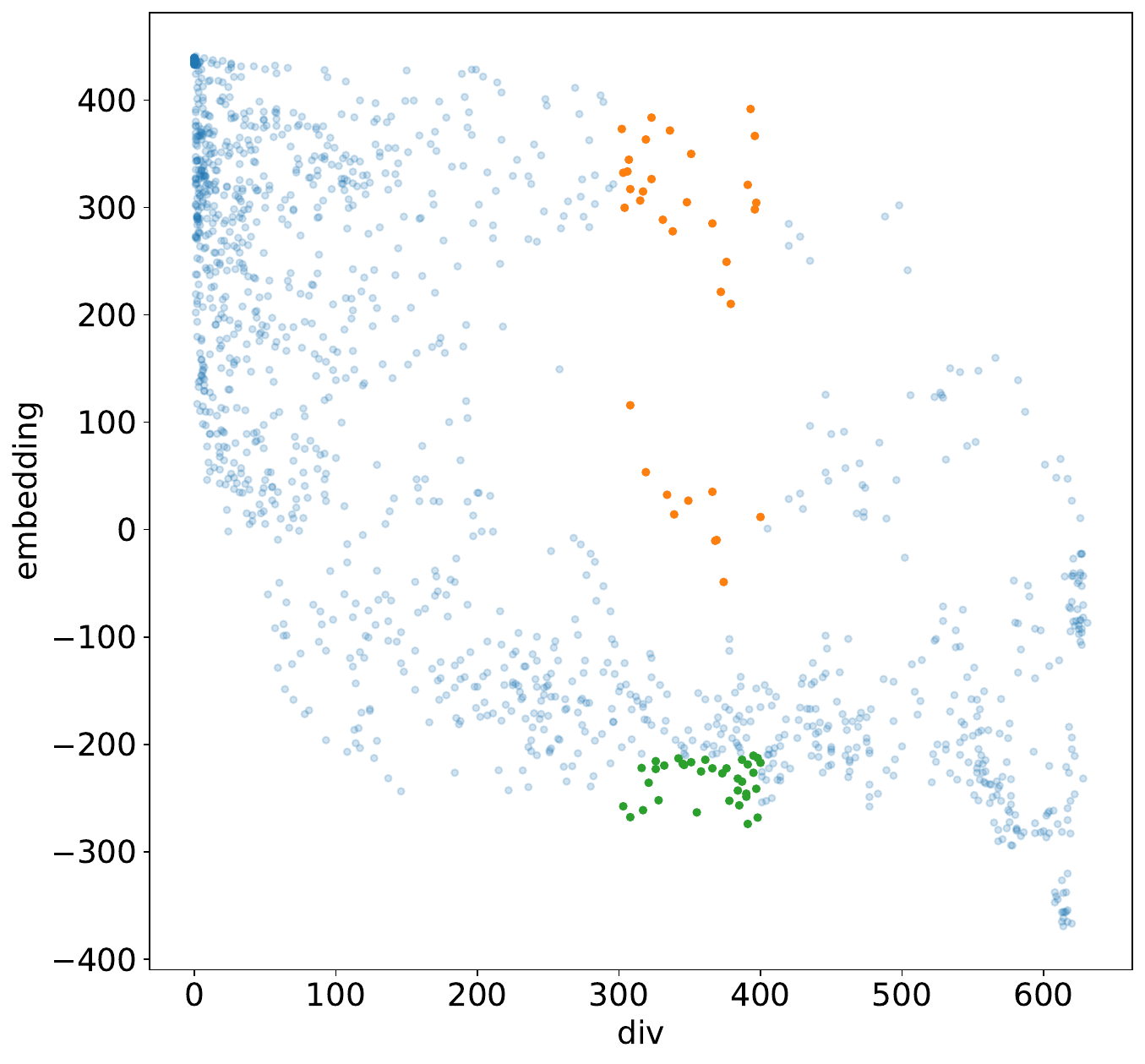}} \\
\subfloat[]%[\emph{}]
   {\includegraphics[width=.38\textwidth]{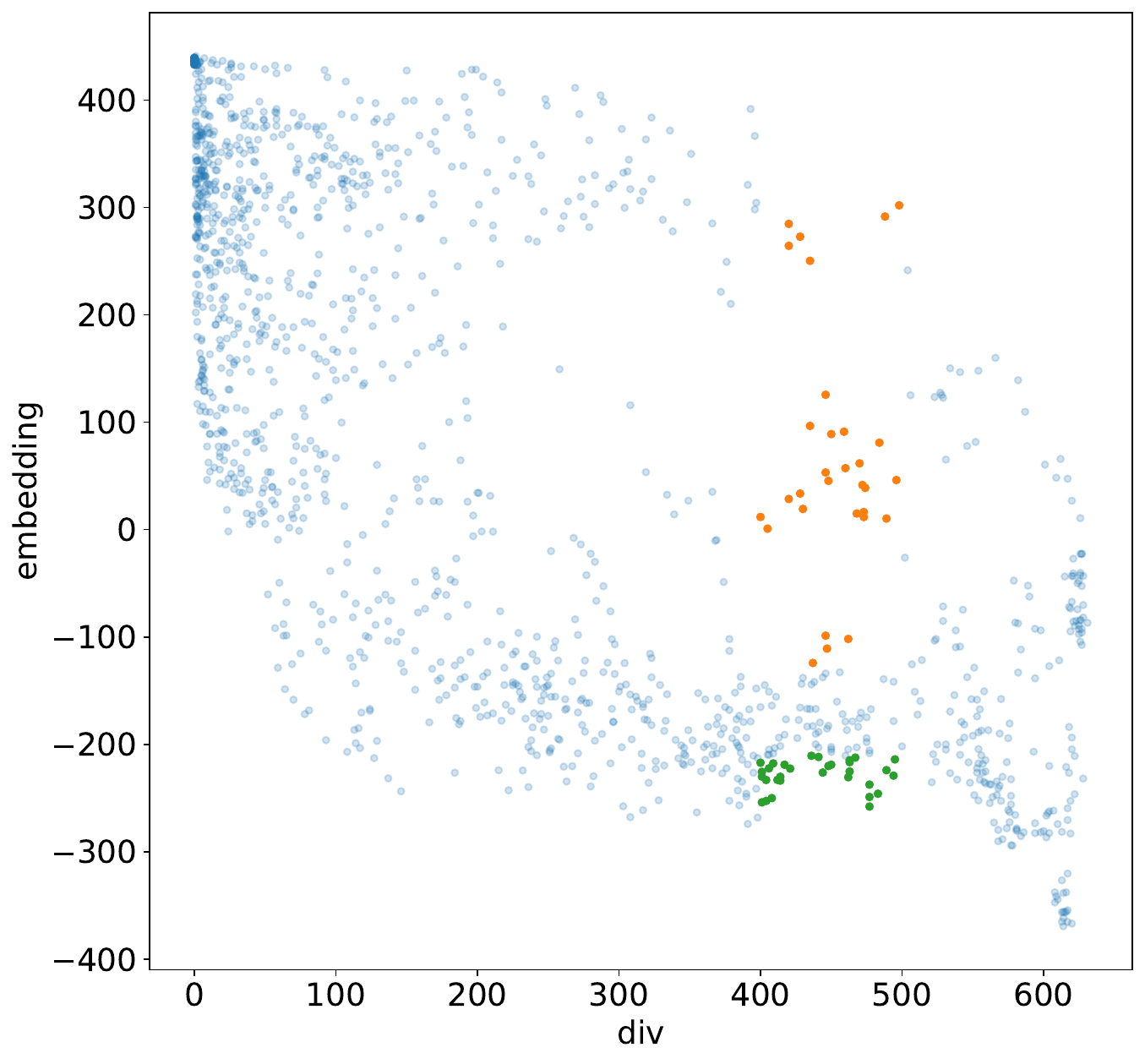}} \quad
\subfloat[]%[\emph{}]
   {\includegraphics[width=.38\textwidth]{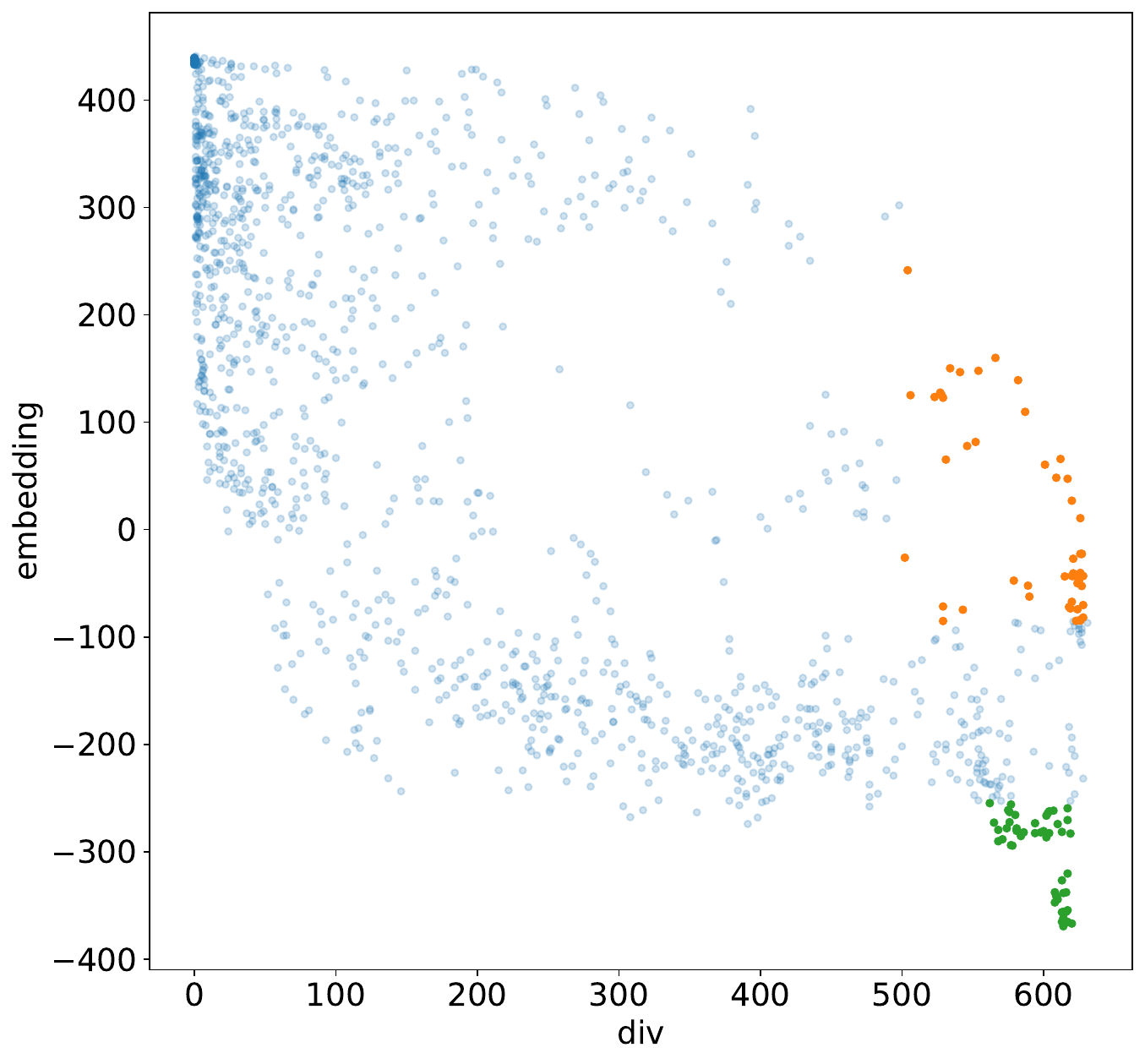}}
\caption{\textbf{Selection procedure of ``upper'' and ``lower'' points}. In each figure, orange points are the 25\% starting from the upper to the lower, and the opposite is true for the greens. For the ``upper'' part, we decided to select the points just below the maximum disease because of the presence of country/year with $0$ technologies.}
\label{fig:selecting}
\end{figure}

\begin{figure}
\centering
\subfloat[]%[\emph{}]
   {\includegraphics[width=.38\textwidth]{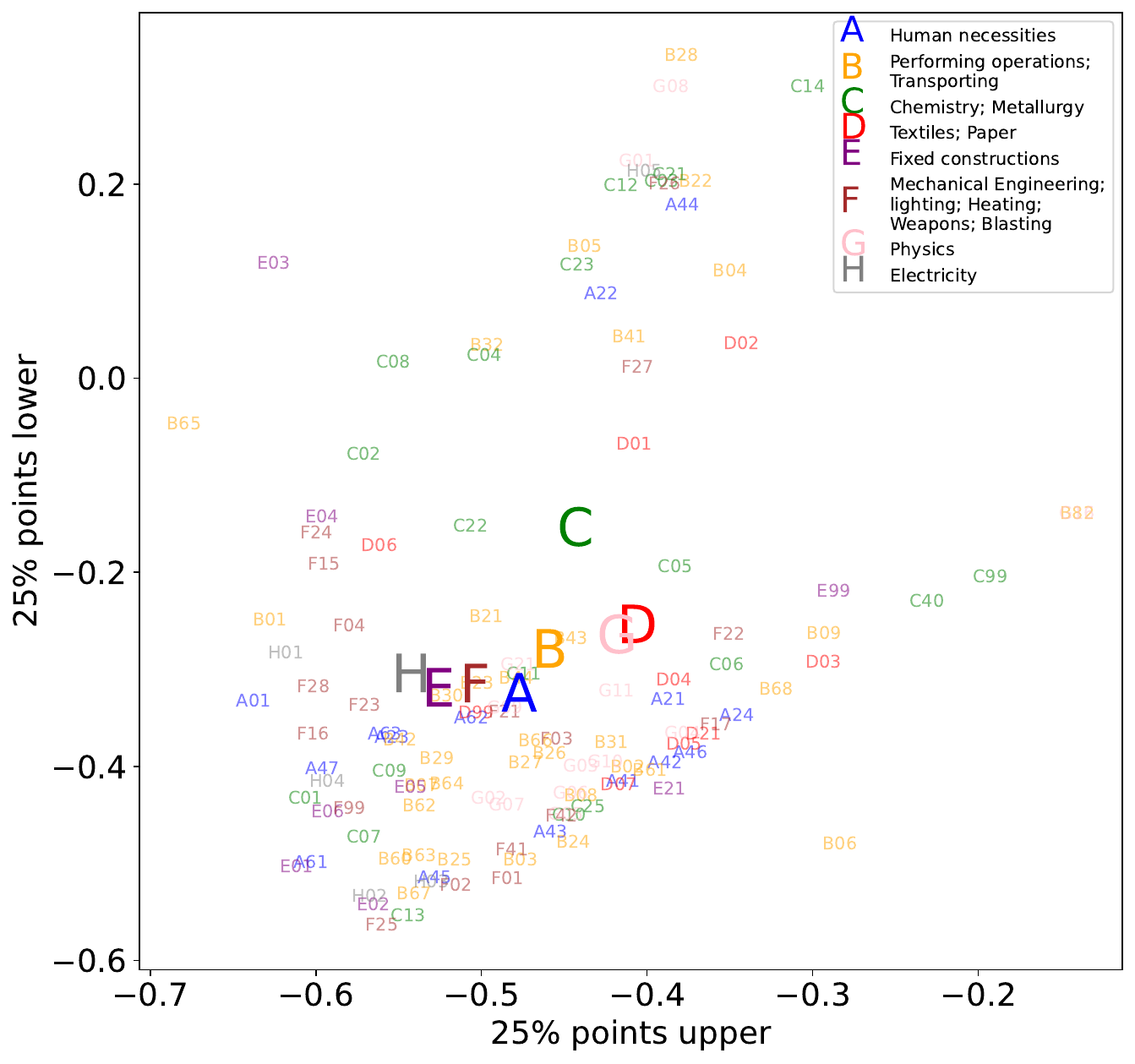}} \quad
\subfloat[]%[\emph{}]
   {\includegraphics[width=.38\textwidth]{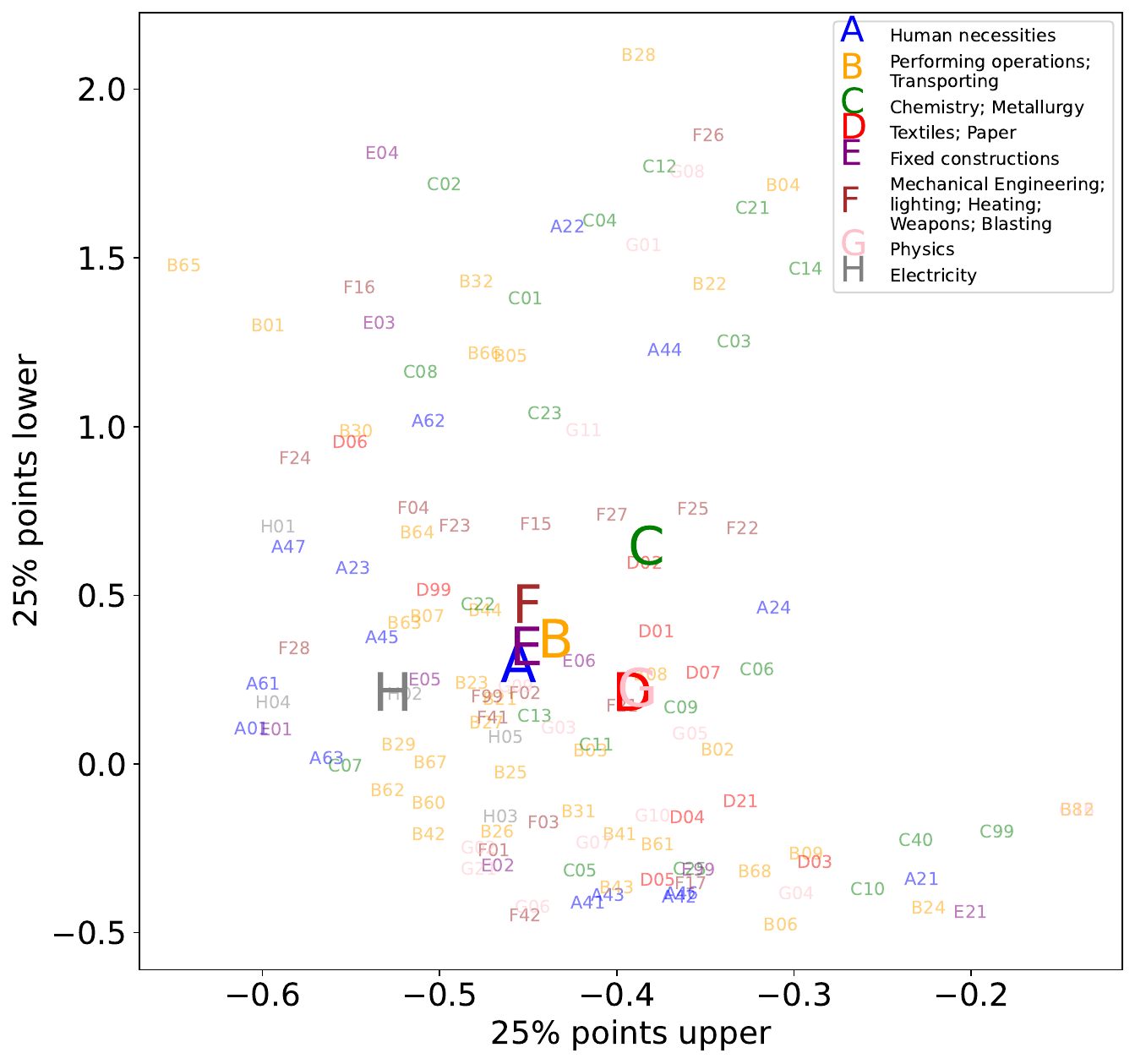}} \\
\subfloat[]%[\emph{}]
   {\includegraphics[width=.38\textwidth]{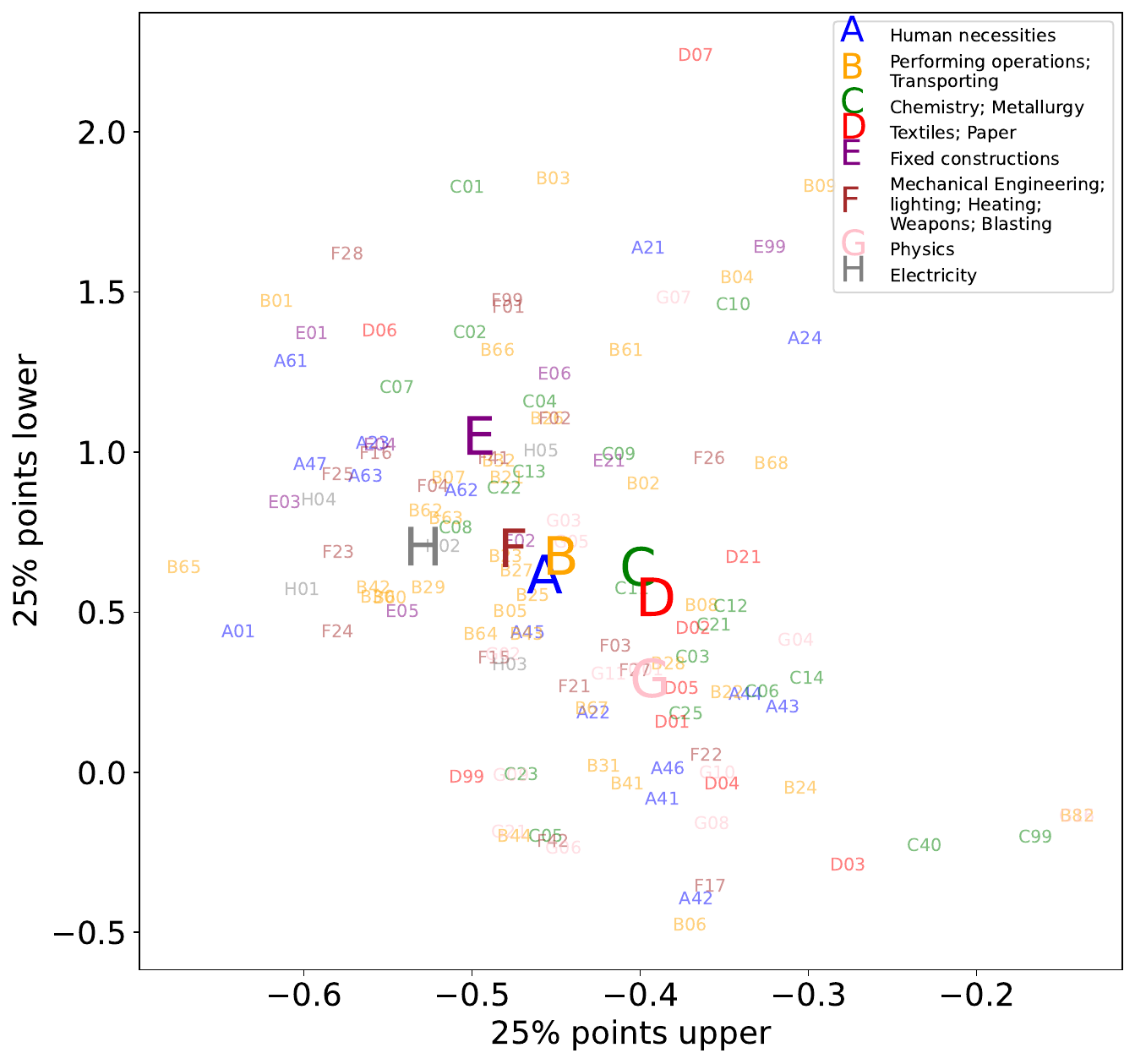}} \quad
\subfloat[]%[\emph{}]
   {\includegraphics[width=.38\textwidth]{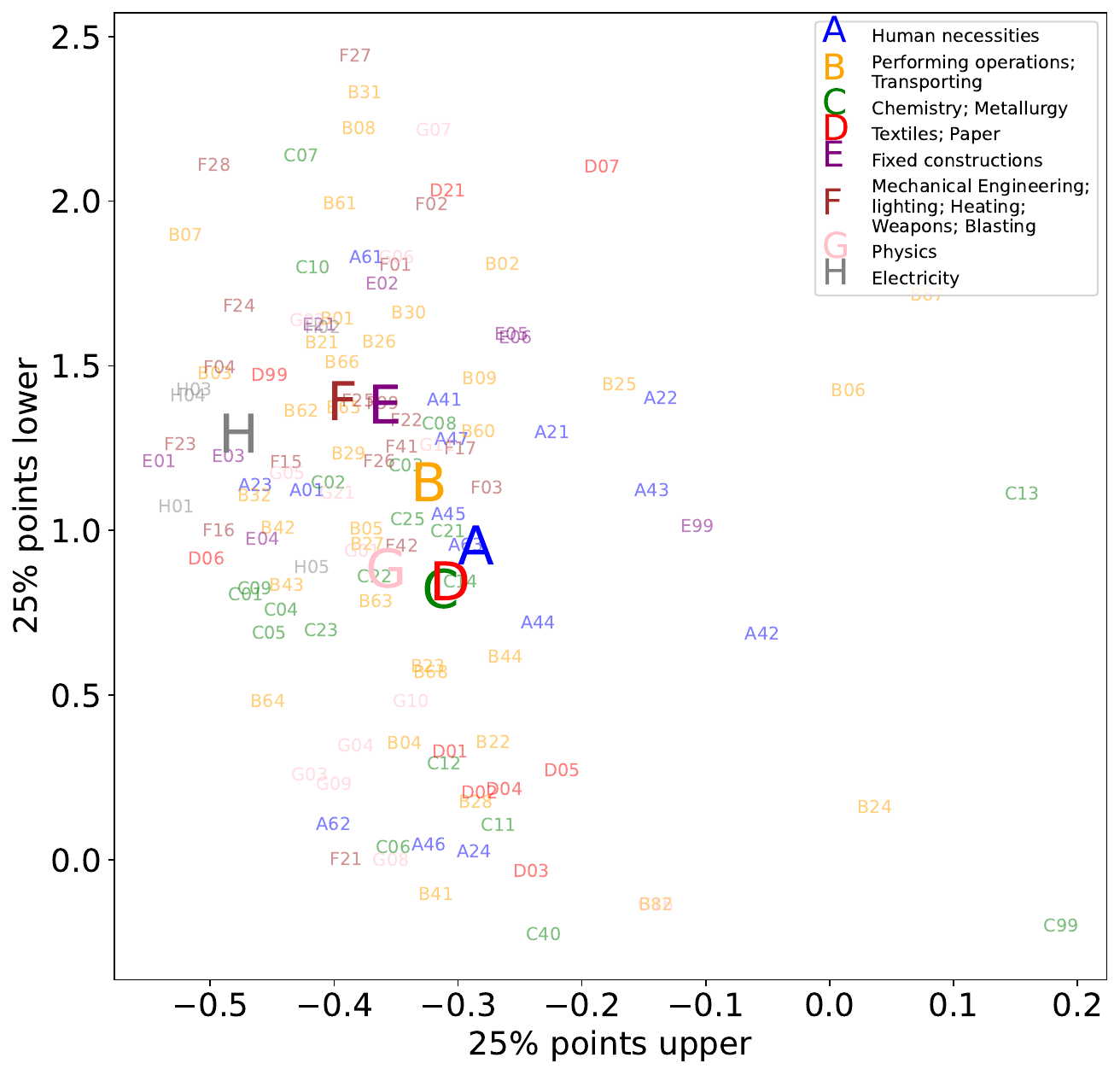}} \\
\subfloat[]%[\emph{}]
   {\includegraphics[width=.38\textwidth]{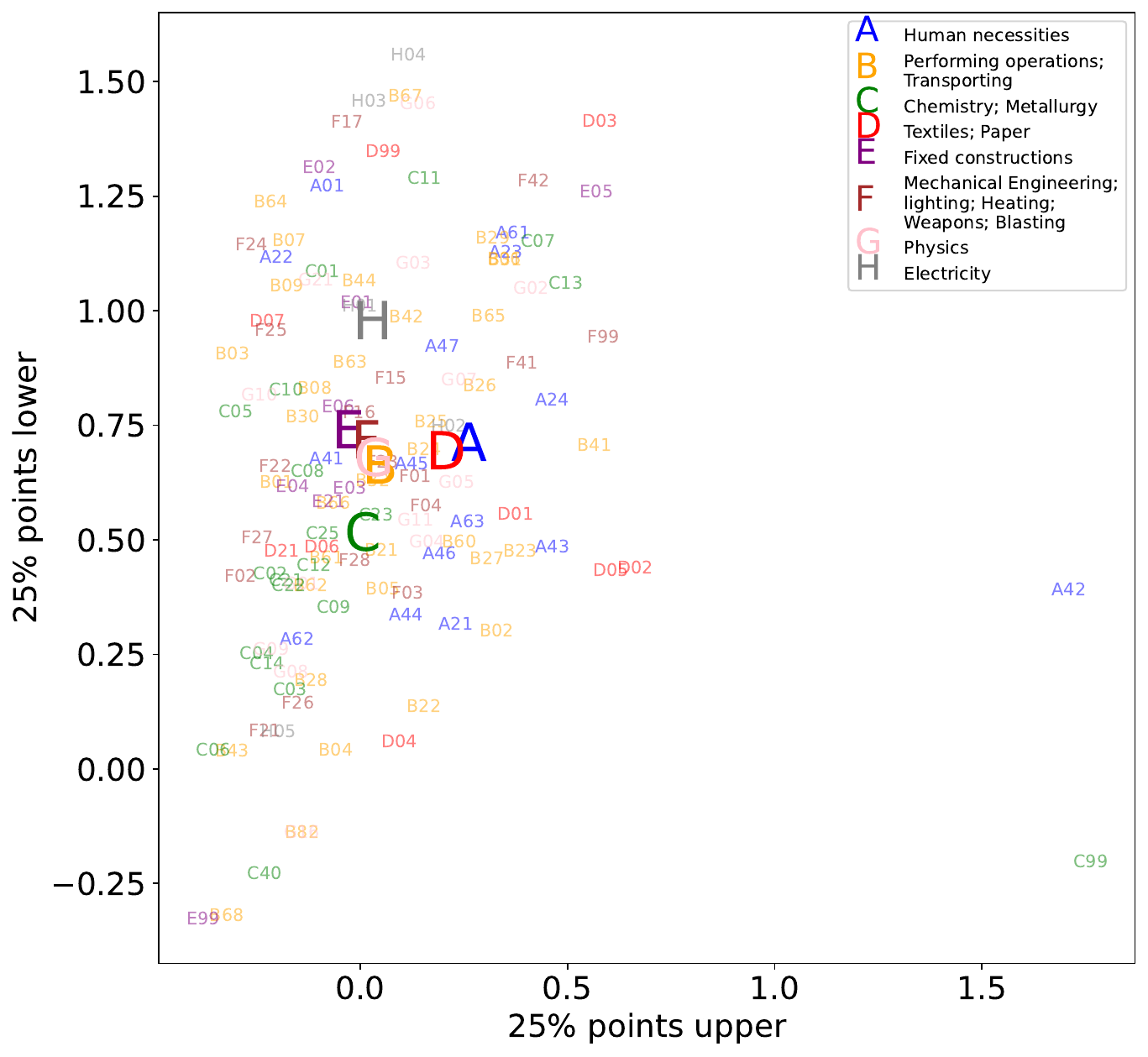}} \quad
\subfloat[]%[\emph{}]
   {\includegraphics[width=.38\textwidth]{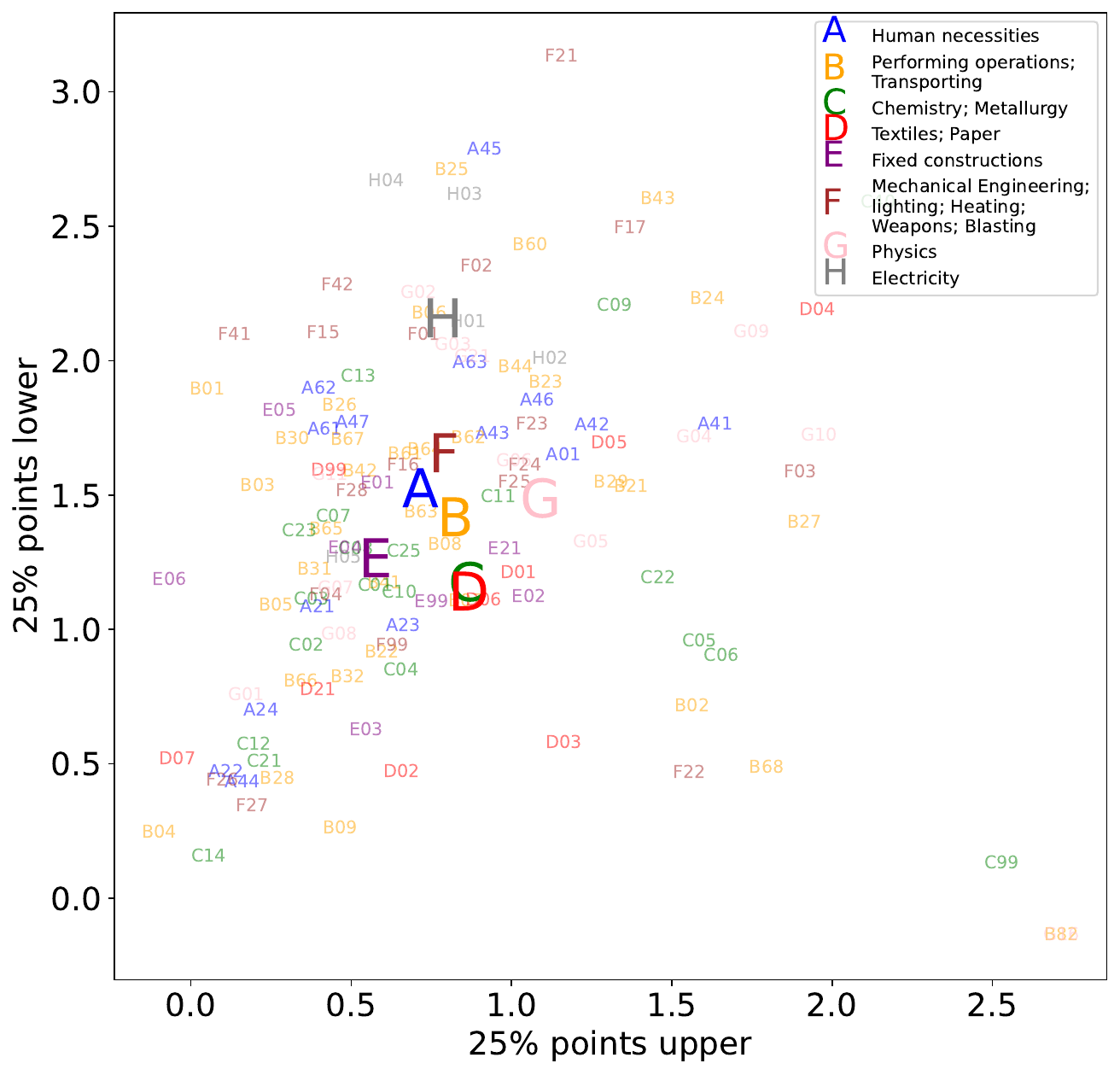}}
\caption{\textbf{Z-score of 3D-technology codes for ``upper'' and ``lower'' parts}. Each text is referred to as a 3D-technology code, and each colour is referred to as one of the eight macrocategory. We compute also the centroid of each 3D code plotting the respective 1D code.}
\label{fig:zscores}
\end{figure}

In this section, we show the intermediate steps used to obtain information about the technology diversification strategies of countries and MAs in the main text.
For each range of diversification of Fig.4, we select the 25\% of ``upper'' and ``lower'' points in both countries and MAs case. We show as an example the country case in Fig\ref{fig:selecting}. For each group, we calculate the z-score for each 3D-technology code to highlight the technology specialization of the two groups. In Fig.\ref{fig:zscores}, we plot the z-scores of ``upper'' and ``lower'' parts with respect to each technology relative to each diversification range of the previous figures. Here, we calculate the centroid for each macrocategory (i.e. 1D-technology). Finally, to highlight the higher differences between the two groups, we select the first 1D technology for each group such that the ranking difference is higher. As example, in Fig\ref{fig:zscores}\textbf{f}, the ranking for the 25\% ``upper'' points is G:8, D:7, C:6, B:5, F:4, H:3, A:2, E:1; instead for the 25\% ``lower'' points is H:1, F:2, A:3, G:4, B:5, E:6, C:7, D:8. The relative ranking distances between the eight macrocategories is, from A to H: -4, 1, 4, 6, -2, -3, 3, -5. This means that the ``upper'' part is more distinct with respect to the ``lower'' one in the D and the lower one in the H. It's important to note that the z-scores of the ``lower'' part are higher than the ``upper'' one; in fact, the ``lower'' countries are more technologically advanced. Here, however, we want to show the technological differences that the different parts focus on in relation to each other to better highlight the technological paths that distinguish the two parts.

\bibliography{sample}